\documentclass[11pt]{article}

\usepackage{graphicx}
\usepackage{latexsym}
\usepackage{hyperref}
\usepackage{hyperref}
\hypersetup{
    colorlinks = true,
    allcolors =blue
}
\usepackage{amsmath}
\usepackage{bm}
\usepackage[numbers]{natbib}
\usepackage{epsfig}
\usepackage{fancyhdr}
\usepackage{sectsty,lscape,url}
\usepackage[all=normal,floats,lists,sections,charwidths]{savetrees}
\usepackage[hmargin=0.8in,vmargin=1in]{geometry}

\usepackage{color}
\newlength{\intwidth}

\def\XXint#1#2#3{{\setbox0=\hbox{$#1{#2#3}{\int}$}
\vcenter{\hbox{$#2#3$}}\kern-.5\wd0}}

\newcommand{\tr}{\ensuremath{\mathrm{tr}}}

\newcommand{\bPsi}{\ensuremath{\bm{\Psi}}}

\newcommand{\bg}{\ensuremath{\mathbf{g}}}

\newcommand{\bk}{\ensuremath{\mathbf{k}}}

\newcommand{\bn}{\ensuremath{\mathbf{n}}}

\newcommand{\br}{\ensuremath{\mathbf{r}}}
\newcommand{\bs}{\ensuremath{\mathbf{s}}}

\newcommand{\bu}{\ensuremath{\mathbf{u}}}

\newcommand{\bx}{\ensuremath{\mathbf{x}}}

\newcommand{\bD}{\ensuremath{\mathbf{D}}}
\newcommand{\bE}{\ensuremath{\mathbf{E}}}

\newcommand{\bG}{\ensuremath{\mathbf{G}}}

\newcommand{\bI}{\ensuremath{\mathbf{I}}}
\newcommand{\bJ}{\ensuremath{\mathbf{J}}}
\newcommand{\bK}{\ensuremath{\mathbf{K}}}
\newcommand{\bL}{\ensuremath{\mathbf{L}}}
\newcommand{\bM}{\ensuremath{\mathbf{M}}}
\newcommand{\bN}{\ensuremath{\mathbf{N}}}

\newcommand{\bR}{\ensuremath{\mathbf{R}}}

\newcommand{\bU}{\ensuremath{\mathbf{U}}}

\newcommand{\bW}{\ensuremath{\mathbf{W}}}

\newcommand{\sgn}{\ensuremath{\mathrm{sgn}}}

\newcommand{\rd}{\ensuremath{\mathrm{d}}}

\newcommand{\gn}{\ensuremath{\nu}}
\newcommand{\gs}{\ensuremath{\sigma}}

\begin{document}
\hrule
\begin{center}
{\textbf{\Large{Kinematics of a Fluid Ellipse in a Linear Flow}}}\\
\vspace{2mm}

{Jonathan M. Lilly}\\\vspace{0.5mm}
NorthWest Research Associates, eponym@jmlilly.net
\vspace{-2mm}
\end{center}

\begin{abstract}
A four-parameter kinematic model for the position of a fluid parcel in a time-varying ellipse is introduced. For any ellipse advected by an arbitrary linear two-dimensional flow, the rates of change of the ellipse parameters are uniquely determined by the four parameters of the velocity gradient matrix, and vice versa. This result, termed \emph{ellipse/flow equivalence}, provides a stronger version of the well-known result that a linear velocity field maps an ellipse into another ellipse. Moreover, ellipse/flow equivalence is shown to be a manifestation of Stokes' theorem. This is done by deriving a matrix-valued extension of the classical Stokes' theorem that involves a spatial integral over the velocity gradient tensor, thus accounting for the two strain terms in addition to the divergence and vorticity. General expressions for various physical properties of an elliptical ring of fluid are also derived. The ellipse kinetic energy is found to be composed of three portions, associated respectively with the circulation, the rate of change of the moment of inertia, and the \emph{variance} of parcel angular velocity around the ellipse. A particular innovation is the use of four matrices, termed~the $\bI\bJ\bK\bL$ basis, that greatly facilitate the required calculations. 
\end{abstract}

\begin{center}
Final author's copy, in press at \emph{Fluids}, \url{http://www.mdpi.com/journal/fluids}, 2018.
\end{center}

\hrule\vspace{2mm}

\section{Introduction}
\thispagestyle{empty}

Elliptical vortex solutions form a fundamental building block of fluid dynamics. In two-dimensional flow, the Kida vortex (an elliptical vortex patch evolving under the action of a linear flow field) is~one of the few exact solutions of the Euler equations. This solution was introduced by Kida \cite{kida81-jpsj} as an~extension of the classical unforced Kirchhoff vortex, given a Hamiltonian formulation by Neu~\cite{neu84-pf}, then generalized to time-dependent forcing fields by Ide and Wiggins \cite{ide95-fdr}. The Kida vortex has been of considerable interest as a means of understanding such phenomena as instability mechanisms \cite{dritschel90-jfm,meacham90-dao,bayly96-ptrsla,mitchell08-pf,guha13-pre,koshel17-npg}, chaotic advection \cite{bertozzi88-sjma,polvani90-science}, the interaction of diffusion and advection \cite{koshel13-npg}, vortex-vortex interactions in shear \cite{ngan96-pf}, vortex energetics \cite{vanneste10-pf} and vortex interactions with boundaries \cite{crosby13-pf}. Elliptical vortices also play a central role as a basis ingredient in ambitious attempts to approximate the dynamics of more complex or realistic flows \cite{melander86-jfm,legras91-pfa,dritschel91-pfa,meacham97-pf}. Closely related to such elliptical vortices are \emph{ellipsoidal} vortices in three dimensions, which have been studied under quasi-geostrophic dynamics \cite{meacham92-dao,mckiver03-jfm,mckiver06-jfm,mckiver15-amp}. Elliptical~vortices are also relevant for two-dimensional surface quasi-geostrophic dynamics, in which a steadily-rotating, unforced elliptical vortex solution has been found \cite{dritschel11-gafd}, consisting of a non-uniform distribution of surface buoyancy having a particular form.
 
Elliptical vortex solutions are important in shallow water dynamics, as well. A freely-evolving elliptical vortex within an active layer that outcrops at the surface is thought to be a reasonable model for large-scale oceanic eddies such as Gulf Stream rings (e.g., \cite{cushman-roisin85-jgr,young86-jfm}). Such shallow water eddies may exhibit a range of behaviors. Two exact analytic solutions were found by Cushman-Roisin and collaborators \cite{cushman-roisin85-jgr,cushman-roisin87-tellus}: a \emph{rodon} or freely-precessing elliptical eddy, the shallow-water analog of a~Kirchhoff vortex; and a \emph{pulson} (a term coined by \cite{kirwan91-ntop}), a circular eddy with a moment of inertia that pulsates at the inertial frequency. These two solutions were later combined by \citet{rogers89-pla} to give the \emph{pulsrodon}. A final degree of freedom, corresponding to a time-varying eccentricity or elliptical aspect ratio, was examined by both \citet{young86-jfm} and \citet{holm91-jfm}, drawing on earlier work by \citet{ball63-jfm} on fluid motion in a parabolic basin; while no analytic solution exists in this case, there is a simple differential equation that governs the evolution of the aspect ratio. On a more practical level, the differing stabilities of cyclonic and anticyclonic vortices in shallow water dynamics has been implicated as a possible explanation for the well-known dominance of anticyclones in the world ocean \cite{arai94-chaos,stegner00-jpo}.



Elliptical vortex solutions would appear to be of renewed relevance for oceanography on account of the groundbreaking recognition of the ubiquity of propagating nonlinear vortices in satellite altimetry by Chelton et al. \cite{chelton11-pio}. While such structures appear circular when seen with the limited resolution available to remotely-sensed sea surface height, high-resolution numerical models such as that of Early et al. \cite{early11-jpo} reveal them to be approximately elliptical in shape; see Figure~7 therein. One particular application for analytic solutions to elliptical vortices is as test cases for assessing Lagrangian analysis methods. For example, Lilly et al. \cite{lilly11-grl} present an objective method for inferring time-varying properties of a possibly elliptical vortex, by decomposing a Lagrangian trajectory into a quasi-periodic or oscillatory portion and a residual. Analytic solutions could be used to validate such methods and to determine to what extent, and under what conditions, elliptical structure in oceanic eddies may be accurately inferred on the basis of Lagrangian observations. Yet, the usefulness of such analytic solutions is limited by the dispersed state in which the relevant results appear throughout the literature, a difficulty that is compounded by the formidable algebra that is often required. 


The goal of this paper is to set the stage for a unified exploration of elliptical vortex solutions, by establishing kinematic results that are common to both the two-dimensional and shallow water systems. All of the solutions referred to above have in common the fact that they consist of elliptical rings of fluid advected by a constant or time-varying linear flow field. In fact, the bulk of the mathematical machinery that is required for treating elliptical vortex solutions is connected not with any characteristic of the particular physical systems, but rather with the underlying kinematics of deformable fluid ellipses. The~creation of a common mathematical framework allows the separation of the mathematical and physical aspects of the investigation, permitting solutions to be derived with much greater ease. The~key is a parametric model for the position of a fluid parcel within a time-varying ellipse, controlled by three parameters describing the ellipse geometry, together with a fourth describing the parcel location around the ellipse periphery. A matrix-based approach is employed that greatly simplifies algebraic manipulations and that, it is hoped, may find applicability in other areas, as well. The focus of this paper is therefore abstract in nature; the actual derivation and examination of the vortex solutions is outside the scope of this work and will be left to a sequel. 

It is well known (see, e.g., \cite{kida81-jpsj}) that a two-dimensional linear velocity field maps an ellipse into another ellipse. Provided that the positions of particles along the ellipse are tracked in addition to the ellipse geometry, the evolution of a fluid ellipse is even more directly related to the linear flow field. They are equivalent: a linear flow determines the evolution of any ellipse placed within it, and~conversely, the evolution of any (non-degenerate) fluid ellipse uniquely specifies the linear flow that must have created it. In an exact sense, the evolution of a fluid ellipse is the Lagrangian representation of a linear flow. This result, which will be termed \emph{ellipse/flow equivalence}, will be shown to be a special case of a generalization of Stokes' theorem to a $2\times2$ matrix form. This extended Stokes' theorem embodies not only the classical Stokes' theorem and the divergence theorem, but also analogous integral relations between spatial and contour integrals of the two components of the strain field. This result is not strictly new, in the sense that it is merely another form of the generalized Stokes' theorem, like the divergence theorem; yet, its presentation here has a strikingly simple form, incorporating all four components of the velocity gradient tensor, that does not appear to have been previously presented. 

Using the kinematic model, general expressions for the basic physical properties of an elliptical ring of fluid are derived, specifically the circulation, the average angular momentum around the ellipse, and the average kinetic energy. These are expressed both in terms of the rates of change of the ellipse itself, as well as in terms of the flow derivatives. While the expressions for the circulation and angular momentum have appeared previously (e.g., \cite{holm91-jfm}), that for the kinetic energy is new. It is shown that in a non-divergent, temporally-constant imposed flow field, the angular momentum averaged along any fluid ellipse, perhaps surprisingly, remains constant. The kinetic energy is shown to have a~partitioning into three terms: a term associated with the circulation, a term associated with the rate of change of the moment of inertia of the elliptical ring, and a third term, combining the effects of both deformation and precession, which is shown to be due to the variance of parcel angular velocity (or momentum) around the ellipse. Rearranging this expression for kinetic energy illustrates the possibility of fixed energy solutions in which oscillations of the moment of inertia decouple from other changes in the ellipse geometry. This purely kinematic result is reminiscent of a theorem of Ball's \cite{ball63-jfm} on the dynamics of the moment of inertia that is central to the study of elliptical vortices in shallow water \cite{young86-jfm,holm91-jfm}.




Herein, only passive rings of fluid are considered, that is, those lacking a vorticity anomaly relative to the background. While this may seem like a stringent limitation, that is not the case. The reason is that in all of the vortex solutions referred to earlier, the self-advection velocity arising from the vorticity anomaly of the elliptical vortex itself is also linear. Most of the results derived herein therefore apply directly to those cases with only minor modifications.


The structure of the paper is as follows. Section~\ref{definitions} introduces definitions and notation, including the matrix basis that will be used herein, and presents the elementary properties of a linear flow; the~kinematic model for a fluid ellipse is introduced in Section~\ref{kinematics} and is used to derive the principle of ellipse/flow equivalence; the integral properties of an elliptical ring of fluid are derived in Section~\ref{integrals}; in~Section~\ref{stokes}, the extended Stokes' theorem is derived, and the ellipse/flow equivalence is reinterpreted as a special case of this theorem. The paper concludes with a discussion.


\section{Definitions and Notation} \label{definitions}

This section begins with a discussion of a linear velocity field, introduces a matrix-based notation for representing the velocity gradient tensor and relevant differential operators, then uses that basis to derive further properties of the linear velocity field. Two sets of Eulerian ellipses characterizing the flow are also examined. 

\subsection{A Linear Velocity Field}

In this paper, we will be primarily concerned with linear velocity fields, that is, velocity fields $\bu(\bx,t)$ that depend linearly on the horizontal position~$\bx$. We may write the velocity in terms of some two-by-two matrix $\bU(t)$ as
\begin{equation}\label{linvel}
\bu(\bx,t)=\bU(t)\bx 
\end{equation}
where the velocity and position vectors have components $ \bu=\left[ u \,\, v\right]^T$ and $\bx=\left[x \,\, y\right]^T$, respectively, with~the superscript ``$T$'' denoting the transpose. Taking the gradient of Equation~(\ref{linvel}) leads to
\begin{equation}\label{ugrad}
\nabla \bu= \begin{bmatrix} \frac{\partial }{\partial x} u & \frac{\partial }{\partial y}u \vspace{.02in}\\ \frac{\partial }{\partial x} v & \frac{\partial }{\partial y}v \end{bmatrix} =\bU
\end{equation} 
and thus the matrix $\bU(t)$ is seen to be identical to the velocity gradient matrix of the linear flow. While~$\nabla \bu$ in general depends on the position $\bx$, for linear flow it becomes independent of $\bx$, and~in this case we write it as $\bU(t)$. We refer to $\bU(t)$ simply as the \emph{flow matrix}, a term already introduced by~\cite{mckiver03-jfm}. Any two-dimensional flow matrix may be written as 
\begin{equation}\label{Udecomp}
\bU(t) =\frac{1}{2}\delta
 \begin{bmatrix}
  1 & 0 \\ 0 & 1
\end{bmatrix}+\frac{1}{2}\zeta\begin{bmatrix}
  0 & -1 \\ 1 & 0
\end{bmatrix}
+\frac{1}{2}\gamma \begin{bmatrix}
  \cos2 \alpha & \sin2 \alpha \\ \sin2 \alpha & -\cos2 \alpha
\end{bmatrix}
\end{equation}
where $\delta$ is the divergence, $\zeta$ is the vorticity, $\gamma$ is the strain magnitude, and $\alpha$ is the strain orientation, all~of which are spatially uniform but potentially time-varying. 

In what follows, it will prove convenient to express differential operators in a matrix-based notation. We introduce
the counterclockwise rotation matrix through angle $\theta$, and the ninety-degree counterclockwise rotation matrix, respectively, as\begin{equation}
\bR(\theta)\equiv \begin{bmatrix}
  \cos\theta &-\sin \theta\\ \sin \theta & \cos\theta
\end{bmatrix},\quad\quad
\bJ \equiv\begin{bmatrix}
  0 &-1\\ 1 & 0
\end{bmatrix}=\bR(\pi/2).
\end{equation}
Furthermore, let boldface $\bm\nabla$ be the horizontal gradient operator represented as a two-vector, $\bm\nabla\equiv\left[ \frac{\partial}{\partial x} \,\,\, \frac{\partial}{\partial y}\right]^T$, as opposed to the basis-free representation $\nabla$. In this notation, the velocity gradient matrix is given by $\nabla \bu= \left(\bm\nabla \bu^T\right)^T$, with the extra transpose required in order that the operator $\bm\nabla$ acts from the left and at the same time recovering the correct arrangement of terms as seen in Equation~(\ref{ugrad}). The velocity divergence and the vertical component of vorticity are then 
\begin{equation}\label{divandcurl}
\delta\equiv\bm\nabla^T \bu = \nabla \cdot \bu = \frac{\partial u}{\partial x}+\frac{\partial v}{\partial y},\quad\quad
\zeta\equiv\bm\nabla^T\bJ^T \bu = \bk\cdot\nabla\times \bu =\frac{\partial v}{\partial x}-\frac{\partial u}{\partial y}
\end{equation}
where $\bk$ is the vertical unit vector. The normal strain $\gn$, shear strain $\gs$, strain magnitude $\gamma$, and strain angle $\alpha$ are defined as
\begin{equation}
\gn \equiv \frac{\partial u}{\partial x}-\frac{\partial v}{\partial y},\quad\quad \gs \equiv
\frac{\partial v}{\partial x}+\frac{\partial u}{\partial y},\quad\quad\gamma\equiv\sqrt{\gn^2+\gs^2},\quad\quad \alpha \equiv \frac{1}{2}\arctan(\gs/\gn)
\end{equation}
while the rate-of-strain matrix for a strain field oriented along direction $\alpha$ arises from the rotation
\begin{equation}\label{strainmatrix}
\frac{1}{2}\bR(\alpha) \begin{bmatrix}
  \gamma & 0 \\ 0 & -\gamma 
\end{bmatrix} \bR^T(\alpha) = \frac{1}{2}\gamma \begin{bmatrix}
  \cos2 \alpha & \sin2 \alpha \\ \sin2 \alpha & -\cos2 \alpha
\end{bmatrix} = \frac{1}{2} \begin{bmatrix}
  \gn & \gs \\ \gs & -\gn
\end{bmatrix} 
\end{equation}
which is the third quantity appearing in Equation~(\ref{Udecomp}). Note that the orientation angle $\alpha$ is defined such that $\alpha=0$ corresponds to extension along the $x$-axis and compression along the $y$-axis. Sometimes, it will be useful to think of the strain in terms of its magnitude $\gamma$ and angle $\alpha$, and other times in terms of the normal strain and shear strain components, $\gn$ and $\gs$.

For future reference, we also note some expressions involving the cross product. If $\mathbf{f}$ and $\bg$ are both purely horizontal vectors, e.g., $\mathbf{f}= \left[ f_x \,\, f_y\right]^T$ and $\mathbf{g}=\left[ g_x \,\, g_y\right]^T$, we have
\begin{equation}
\bk\cdot \left(\mathbf{f}\times \bg\right) =\mathbf{f}^T\bJ^T\mathbf{g}\label{crossproduct},\quad\quad
\bk\times \mathbf{f} = \bJ\mathbf{f}
\end{equation}
for the vertical component of their cross product and for the cross product of the vertical unit vector with the purely horizontal vector $\mathbf{f}$, respectively. In these expressions, $\mathbf{f}$ and $\bg$ are regarded on the left-hand sides as being three-vectors with a vanishing $z$ component. 

\subsection{A Matrix Basis}
 
\begin{figure}[t]
\begin{center}
\includegraphics[width=6in,angle=0]{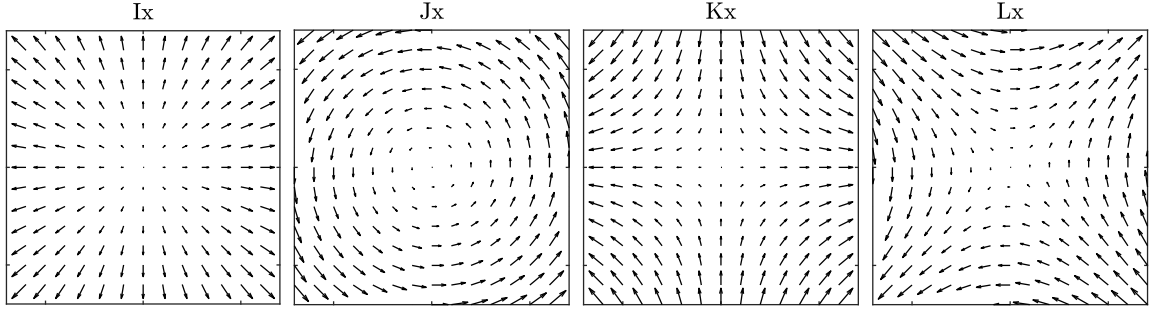}
\end{center}
	\caption{Matrix products associated with the $\bI\bJ\bK\bL$ matrices presented as quiver plots. From left to right, plots of $\bI\bx$, $\bJ\bx$, $\bK\bx$, and $\bL\bx$ are shown. These are the same as a velocity field of pure divergence, pure vorticity, pure normal strain and pure shear strain, respectively. }\label{vectors}
\end{figure}

Any real-valued two-by-two matrix can be written as a weighted sum of the four matrices
\begin{equation}\label{basisdef}
\bI \equiv \begin{bmatrix} 1 & 0 \\ 0 &1\end{bmatrix},\quad\quad
\bJ \equiv \begin{bmatrix} 0 & -1 \\ 1 & 0\end{bmatrix},\quad\quad
\bK \equiv \begin{bmatrix} 1 & 0 \\ 0 &-1\end{bmatrix},\quad\quad
\bL \equiv \begin{bmatrix} 0 & 1 \\ 1 & 0\end{bmatrix}
\end{equation}
which will be referred to as the $\bI\bJ\bK\bL$ basis. The trace, or sum of the diagonal elements, is equal to two~for the identity matrix $\bI$ and zero for the others, while the determinants are positive one for $\bI$ and $\bJ$, and negative one for $\bK$ and $\bL$. The third and fourth matrices, $\bK$ and $\bL$, are reflection matrices about the lines $y=0$ and $x=y$, respectively. In terms of the $\bI\bJ\bK\bL$ basis, we find
\begin{equation}\label{Udecomp2}
\bU(t) =
\frac{1}{2}\left(\delta \bI + \zeta \bJ+\gn\bK +\gs\bL\right)
\end{equation}
as a compact expression for the flow matrix, which is seen to be identical to Equation~(\ref{Udecomp}). The velocity field is then composed of terms proportional to $\bI\bx$, $\bJ\bx$, $\bK\bx$, and $\bL\bx$, respectively, which are shown for illustration purposes in Figure~\ref{vectors}. Note that this is a general decomposition for the matrix product $\bU \bx$ involving any real two-by-two matrix $\bU$.

This basis expansion greatly facilitates matrix multiplications, as one can refer to predetermined rules. $\bJ$, $\bK$, and $\bL$ multiply themselves according to the rule $-\bJ\bJ=\bK\bK=\bL\bL=\bI$, which is equivalent to the transposition rules $\bJ=-\bJ^T$, $\bK=\bK^T$, and $\bL=\bL^T$. These matrices multiply each other as
\begin{equation}\label{jkl}
%
\begin{array}{llll}
\bJ\bK=+\bL\quad\quad&\bK\bL=-\bJ\quad\quad &\bL\bJ=+\bK\\
\bK\bJ=-\bL\quad\quad&\bL\bK=+\bJ\quad\quad &\bJ\bL=-\bK
\end{array}
\end{equation}
thus forming a kind of cycle. In the upper line, the matrices are arranged in alphabetical order, if~one considers $\bL$ to be followed again by $\bJ$; only the second, beginning with $\bK$, leads to a minus sign. The~reverse alphabetical order rules on the second line are obtained by transposition. The~multiplication rules are also presented in Table~\ref{ijkltable}, the inspection of which reveals a kind of symmetry. It follows from these rules that the expansion coefficients for a general matrix $\bU$ in terms of the $\bI\bJ\bK\bL$ basis are
\begin{equation}\label{jklexpansion}
\bU =\frac{1}{2}\left\{\tr\left\{\bU\right\}\bI+
\tr\left\{\bU\bJ^T\right\}\bJ+\tr\left\{\bU\bK\right\}\bK+\tr\left\{\bU\bL\right\}\bL\right\}
\end{equation}
where $\tr\left\{\cdot\right\}$ denotes the matrix trace. The factor of one half arises due to the fact that $\tr\left\{\bI\right\}=2$.


\begin{table}[b]
\caption{Multiplication rules for the $\bI\bJ\bK\bL$ matrices, giving the result of multiplying the row matrix by the column matrix. For example, the terms in the second row are the values of $\bJ^T\bI$, $\bJ^T\bJ$, $\bJ^T\bK$, and $\bJ^T\bL$.}\label{ijkltable}\vspace{.1in}
\centering
\begin{tabular}{l|rrrr}
\hline
   & $\bI$ & $ \bJ$ & $\bK$ & $ \bL$ \\\hline
$\bI$ & $ \bI$ & $\bJ$ & $\bK$ & $\bL$ \\
$\bJ^T$ & $-\bJ$ & $\bI$ & $-\bL$ & $\bK$ \\
$\bK$ & $\bK$ & $ -\bL$ & $\bI$ & $-\bJ$ \\
$\bL$ & $\bL$ & $\bK$ & $\bJ$ & $\bI$ \\
\hline
\end{tabular}
\end{table}

The matrices $\bJ$, $\bK$, and $\bL$ transform under rotations as follows. Since rotations in two dimensions commute, $\bR(\theta) \bJ\bR^T(\theta)=\bJ$, and $\bJ$ is unchanged by a rotation. The $\bK$ and $\bL$ matrices transform as
\begin{equation}
\bR(\theta) \bK\bR^T(\theta) = \begin{bmatrix}\cos 2\theta & \sin 2\theta \\ \sin2\theta & -\cos2\theta\end{bmatrix},\quad\quad\bR(\theta) \bL\bR^T(\theta) = \begin{bmatrix}-\sin 2\theta & \cos 2\theta \\ \cos2\theta & \sin2\theta\end{bmatrix}
\end{equation}
which we can readily verify by decomposing the rotation matrix as $\bR(\theta)=\cos\theta \bI +\sin\theta \bJ$. Then
\begin{align}\label{Krot}
\bR(\theta) \bK\bR^T(\theta) &=
\left(\cos\theta\bI+\sin\theta\bJ\right)\bK
\left(\cos\theta\bI-\sin\theta\bJ\right)=\cos2\theta\bK+\sin2\theta\bL
\\
\bR(\theta) \bL\bR^T(\theta) &=\label{Lrot}
\left(\cos\theta\bI+\sin\theta\bJ\right)\bL
\left(\cos\theta\bI-\sin\theta\bJ\right)=\cos2\theta\bL- \sin2\theta \bK
\end{align}
as we see at once through the application of the multiplication rules of Equation~(\ref{jkl}). The first of these has been used in forming the rotated strain matrix in Equation~(\ref{strainmatrix}). 

This matrix basis may be used to derive the form of $\bU(t)$ in a direct way. A gradient matrix may be expanded in the $\bI\bJ\bK\bL$ basis as
\begin{equation}
\left(\bm{\nabla}\bu^T\right)^T = \frac{1}{2} \left\{\label{uijkl}
\left(\bm{\nabla}^T\bu\right)\bI
+\left(\bm{\nabla}^T\bJ^T\bu\right)\bJ
+\left(\bm{\nabla}^T\bK\bu\right)\bK
+\left(\bm{\nabla}^T\bL\bu\right)\bL
\right\} 
\end{equation}
which follows from Equation~(\ref{jklexpansion}) if one observes that $\tr\left\{\left(\bm{\nabla}\bu^T\right)^T\bG^T\right\}=\tr\left\{\bG\bm{\nabla}\bu^T\right\}=\bm{\nabla}^T\bG^T\bu$ for a generic real-valued matrix $\bG$, since $\tr\left\{\bG\bm{\nabla}\bu^T\right\}=G_{11}v_x +G_{12}v_y-G_{21}u_x -G_{22}u_y=\bm{\nabla}^T\bG^T\bu$ from direct calculation.  The coefficients of the $\bI\bJ\bK\bL$ matrices in this expression can be thought of in two different ways. They can be seen as the divergences of the original velocity field $\bu$, the rotated velocity field $\bJ^T\bu$, and the two velocity reflected fields $\bK\bu$ and $\bL\bu$, respectively. Equivalently, we may think of these coefficients as being due to the four operators
\begin{equation}
\bm\nabla = \begin{bmatrix} \frac{\partial}{\partial x}\vspace{.03in} \\ \frac{\partial}{\partial y} \end{bmatrix},\quad\quad
\bJ\bm\nabla = \begin{bmatrix} -\frac{\partial}{\partial y} \vspace{.03in} \\ \frac{\partial}{\partial x} \end{bmatrix},\quad\quad
\bK\bm\nabla = \begin{bmatrix} \frac{\partial}{\partial x} \vspace{.03in} \\ -\frac{\partial}{\partial y} \end{bmatrix},\quad\quad
\bL\bm\nabla = \begin{bmatrix} \frac{\partial}{\partial y} \vspace{.03in} \\ \frac{\partial}{\partial x} \end{bmatrix}\label{operatordef}
\end{equation}
acting on the original velocity field $\bu$, by grouping the matrix with $\bm\nabla$ as in $\bm{\nabla}^T\bG^T\bu=\left(\bG\bm{\nabla}\right)^T\bu$. The~first two such operators in Equation~(\ref{operatordef}) are recognized from Equation~(\ref{divandcurl}) as the divergence and~the vertical component of the curl, respectively. Writing out all four quantities explicitly leads to 
\begin{equation}
\left(\bm{\nabla}\bu^T\right)^T = \frac{1}{2} \left\{
\left(\frac{\partial u}{\partial x}+\frac{\partial v}{\partial y}\right)\bI
+\left(\frac{\partial v}{\partial x}-\frac{\partial u}{\partial y}\right)\bJ
+\left(\frac{\partial u}{\partial x}-\frac{\partial v}{\partial y}\right)\bK
+\left(\frac{\partial v}{\partial x}+\frac{\partial u}{\partial y}\right)\bL
\right\}\label{nablauexpand}
\end{equation}
from which we see that the terms in the velocity gradient matrix of Equation~(\ref{ugrad}) are indeed correctly recovered. If the velocity gradient is spatially uniform, then $\left(\bm{\nabla}\bu^T\right)^T =\bU$, and Equation~(\ref{Udecomp2}) follows from the definitions of $\delta$, $\zeta$, $\gn$, and $\gs$. Thus, the four operators in Equation~(\ref{operatordef}) are seen to be those which operate on $\bu$ to give, respectively, the values of the four velocity components shown in Figure~\ref{vectors}.

This matrix basis, which was used previously (with different notation) in \citet{waterman15-jpo}, has been independently introduced by \cite{anstey17-om} and no doubt elsewhere. It is similar to the well-known Pauli basis for $2\times 2$ Hermitian matrices used in quantum mechanics, which being complex-valued, is~not appropriate for the present application. 

\subsection{The Kinetic Energy, Stream Function. and Angular Velocity}

Next, we obtain a simple expression for the kinetic energy. In a linear flow the local kinetic energy at each point is 
\begin{equation}
\frac{1}{2}\|\bu(\bx,t)\|^2 = \frac{1}{2}\bx^T\left(\bU^T\bU\right)\bx.
\end{equation}

\begin{figure}[t]
\begin{center}
\includegraphics[width=6in,angle=0]{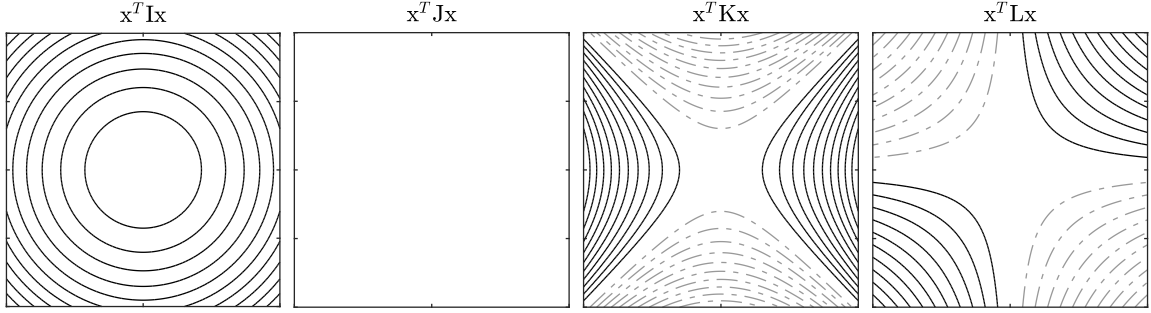}
\end{center}\vspace{-12pt}

	\caption{Quadratic forms associated with the $\bI\bJ\bK\bL$ matrices. From left to right, contour plots of $\bx^T\bI\bx$, $\bx^T\bJ\bx$, $\bx^T\bK\bx$, and $\bx^T\bL\bx$ are shown, with positive contours shown as black solid lines and negative contours as dashed-dotted gray lines. Note that $\bx^T\bJ\bx=0$ identically.}\label{quadratic}
\end{figure}

One may readily find for $\bU^T\bU$, from Equation~(\ref{Udecomp2}) for $\bU$ and using the basis multiplication rules,  
\begin{equation}\label{kematrix}
\bU^T\bU= \frac{1}{4} \left\{\left(\delta^2+\zeta^2+\gamma^2\right)\bI + 2\left(\delta \gn + \zeta \gs \right)\bK + 2\left(\delta \gs-\zeta\gn\right)\bL\right\}
\end{equation}
a form that, it will be seen, indicates that the kinetic energy is constant along either ellipses or hyperbolas.  This equation may be derived by writing $\bU=U_I\bI+U_J\bJ+U_K\bK+U_L\bL$ for convenience, from which one finds $\bU^T\bU= \left(U_I^2+U_J^2+U_K^2+U_L^2\right)\bI + 2\left(U_IU_K+U_JU_L\right)\bK + 2\left(U_IU_L-U_JU_K\right)\bL$. Multiplying Equation~(\ref{kematrix}) by $\bx$ on both sides leads to the terms $\bx^T\bI\bx=x^2+y^2$, $\bx^T\bK\bx=x^2-y^2$ and $\bx^T\bL\bx=2xy$. Thus, the first of the quantities in Equation~(\ref{kematrix}) is non-negative, but the second and third will take on either sign depending on the spatial location. These last two quantities reflect cross terms arising from the interaction of the divergence and the vorticity with the strain. For~reference, the four quadratic products of $\bx$ with itself modified by the $\bI\bJ\bK\bL$ matrices are shown in Figure~\ref{quadratic}.

A non-divergent velocity field may be written in terms of a stream function $\psi(\bx,t)$, given by 
\begin{equation}\label{psidef}
\psi(\bx,t) = \bx^T \bPsi(t) \bx + \psi_o(t)
\end{equation}
where $\psi_o(t)$ is a spatially-constant (but potentially time-varying) term that is generally determined by boundary conditions, and where the \emph{stream function matrix} $\bPsi(t)$ may be chosen to be symmetric, $\bPsi=\bPsi^T$. A linear velocity field $\bu$ with vanishing divergence $\delta$ may then be written as
\begin{equation}
\bu(\bx,t) = \bk \times \nabla \psi = \bJ \bm\nabla \psi = \bJ\bm\nabla \left(\bx^T\bPsi \bx\right) =2\bJ\bPsi\bx ,\quad\quad \delta=0
\end{equation}
using Equation~(\ref{crossproduct}) for the cross product of a vertical vector and a horizontal vector, and noting from direct calculation that $\bm\nabla \left(\bx^T\bPsi \bx\right)=\left(\bPsi+\bPsi^T\right)\!\bx=2\bPsi\bx$. Combining this with $\bu=\bU\bx$, we find
\begin{equation}\label{Psidef}
\bm\Psi(t) = \frac{1}{2}\bJ^T\bU=\frac{1}{4}\left(\zeta \bI +\gs\bK -\gn\bL\right),\quad\quad \delta=0
\end{equation}
for the stream function matrix in terms of the flow matrix. The second equality follows from Equation~(\ref{Udecomp2}) for $\bU(t)$ together with the $\bJ\bK\bL$ multiplication rules of Equation~(\ref{jkl}). Note that the stream function matrix is thus simply half of the flow matrix, rotated ninety degrees clockwise. 

For future reference, we will also define a quantity closely related to the stream function. The~angular velocity about the origin of a parcel at any point in a linear flow can be expressed~as
\begin{equation}
\varpi(\bx,t)\equiv\bk\cdot (\bx \times \bu) = \bx^T \bJ^T \bu = \bx^T \bJ^T \bU\bx
\end{equation}
using Equation~(\ref{crossproduct}) for the cross product. Here, a comment should be made about our choice of terminology. ``Angular velocity'' herein will refer to the angular velocity of the parcel at each point in space, as opposed to an angular velocity due to the solid-body rotation of the entire rotating frame. While $\varpi(\bx,t)$ could also be seen as an angular momentum per unit mass, later angular momentum per unit mass will be discussed as a property of an entire elliptical ring. If the flow is non-divergent, comparison with Equation~(\ref{Psidef}) shows that $\varpi=2\bx^T\bPsi\bx$, and we then have
\begin{equation}\label{psiandangular}
\psi(\bx,t)=\bx^T\bPsi\bx +\psi_o = \frac{1}{2} \bx^T \bJ^T \bU\bx+\psi_o = \frac{1}{2}\varpi +\psi_o
\end{equation}
as the relationship between the stream function and angular velocity at each point. 


\subsection{Measures of Ellipse Size and Shape}

At any time $t$, the equation for an ellipse with orientation $\theta(t)$ and semi-major and semi-minor axes $a(t)$ and $b(t)$ is given by $\bx^T\bE(t)\bx=1$, with the \emph{ellipse matrix} $\bE(t)$ defined as
\begin{equation}\label{ellipsematrix}
\bE(t) \equiv \bR(\theta) \begin{bmatrix} a^{-2} & 0 \\ 0 & b^{-2}\end{bmatrix} \bR^T(\theta) 
\end{equation}
which is observed to be symmetric, $\bE=\bE^T$. The ellipse axes are specified such that $a(t)>b(t)>0$. Note that $\bx^T\bE\bx=c^2$ for different choices of the constant $c$ specifies ellipses with the same orientation and aspect ratio, but with axes scaled as $c a(t)$ and $cb(t)$. 

It is convenient to replace $a$ and $b$ with two other quantities, one measuring the ellipse size or amplitude and another measuring the ellipse shape. Different quantities will prove at times to be more natural than others. The four most common combinations of $a$ and $b$ that will be used are 
\begin{equation}
\rho(t) \equiv \sqrt{ab},\quad\quad I(t) \equiv \frac{a^2+b^2}{2},\quad\quad \eta(t)\equiv \frac{a}{b},\quad\quad \mu(t)\equiv \frac{a^2+b^2}{2ab}=\frac{\eta^2+1}{2\eta} = \frac{I}{\rho^2}
\end{equation}
the first two quantifying the ellipse size and the second two the ellipse shape. Here, $\rho$ is the geometric mean radius, while $I$, the mean squared axes length, will be shown to also be the moment of inertia per unit mass of an elliptical ring of fluid. The third quantity, $\eta$, is the ellipse aspect ratio, while the fourth, $\mu$, will be called the \emph{extension}, a name that reflects how it changes as the ellipse is distorted. The extension $\mu$ equals unity for a circle and increases without bound as the aspect ratio $a/b$ increases with the area held fixed. Another interpretation of $\mu$ is as a nondimensional moment of inertia.

Table~\ref{eccentricitytable} compares four different measures of ellipse shape. Three of these, those above the horizontal line, are expressed in terms of all of the others. The second quantity varies between zero for a circle and unity for a line, and is referred to as the ellipse \emph{linearity} by \cite{lilly10-itsp}; it is seen to arise naturally in the context of time series analysis of complex-valued signals. The final quantity is somewhat less fundamental, but will nevertheless appear frequently in what follows. This table is useful in understanding how the appearance of various geometric terms changes depending on one's choice of shape measure. The classical eccentricity, $\epsilon\equiv\sqrt{1-b^2/a^2}$, does not appear to be useful for this problem and is therefore not presented here. One may also note, for future reference, 
\begin{equation} \label{derivrelations}
\frac{\rd\ln \eta}{\rd t} =\frac{1}{\sqrt{\mu^2-1}}\frac{\rd\mu }{\rd t} =\frac{1}{1-\lambda^2}\frac{\rd\lambda}{\rd t}
\end{equation}
as the relationships between the time derivatives of the aspect ratio, extension and linearity. Note that these derivative expressions require $\eta=a/b\ge 1$, as has been assumed. 

\begin{table}[b]
\caption{A comparison of four different quantities describing ellipse shape. Each quantity is expressed in terms of the first three, as well as in terms of the major and minor semi-axis lengths $a$ and $b$. }\vspace{.1in}\label{eccentricitytable}
\centering
\begin{tabular}{lccc|ccccc}
\hline
\textbf{Name} & \textbf{Symbol} &\textbf{Range} 	& \boldmath$(a,b)$	&\boldmath$(\eta)$&\boldmath$(\lambda)$ & \boldmath$(\mu)$ \\
\hline\vspace{.05in}
Aspect ratio &$\eta$& $(1,\infty]$ &	$\displaystyle\frac{a}{b}$			&$\eta$&$\displaystyle\sqrt{\frac{1+\lambda}{1-\lambda}}$ &$\displaystyle\sqrt{\frac{\mu+\sqrt{\mu^2-1}}{\mu-\sqrt{\mu^2-1}}}$\\\vspace{.05in}
Linearity & $\lambda$ &$(0,1)$ 	& $\displaystyle\frac{a^2-b^2}{a^2+b^2}$ & $\displaystyle\frac{\eta^2-1}{\eta^2+1}$ & $\lambda$&$\displaystyle\frac{\sqrt{\mu^2-1}}{\mu}$\\\vspace{.05in}
Extension \!\!\! &$\mu$ &$(1,\infty]$ & $\displaystyle\frac{a^2+b^2}{2ab}$ & $\displaystyle\frac{\eta^2+1}{2\eta}$ & $\displaystyle\frac{1}{\sqrt{1-\lambda^2}}$ & $\mu$\\\hline\vspace{.05in}
--- & $\displaystyle\frac{\lambda}{\mu}$ &$(0,\infty]$ & $\displaystyle\frac{a^2-b^2}{2ab}$ & $\displaystyle\frac{\eta^2-1}{2\eta}$ & $\displaystyle\frac{\lambda}{\sqrt{1-\lambda^2}}$ & $\sqrt{\mu^2-1}$\\
\hline
\end{tabular}
\end{table}

\subsection{Stream Function and Energy Ellipses}


In the next section, we will investigate ellipses composed of sets of Lagrangian particles. Before~examining these Lagrangian ellipses, we now compute two sets of Eulerian ellipses associated with the flow itself. The first type comprises contours of a constant stream function for a non-divergent flow field, and the second type contours of constant kinetic energy. A general matrix $\bE$ of the form
\begin{equation}\label{Eexpansion}
\bE =\bR(\theta) \begin{bmatrix} a^{-2} & 0 \\ 0 & b^{-2}\end{bmatrix} \bR^T(\theta) =
\frac{a^{-2}+b^{-2}}{2}\,\bI -\frac{b^{-2}-a^{-2}}{2}\left(\cos2\theta\bK+\sin2\theta\bL\right)
\end{equation}
specifies a family of concentric ellipses as $\bx^T\bE\bx=c^2$ for some constant $c$; note that the second coefficient on the right-hand side is negative due to the fact that $a>b$. Writing $\bE=E_I\bI + E_K \bK + E_L \bL$, one may readily find that the ellipse parameters $\rho^2$, $\mu$, and $\theta$ are given in terms of the matrix components as 
\begin{equation}
\rho^2 = \frac{1}{\sqrt{\det\{ \bE\}\,}},\quad\quad \mu = \frac{1}{2}\frac{\tr\{ \bE\}\,}{\sqrt{\det\{ \bE\}}},\quad\quad\theta = \frac{1}{2} 
 \arctan\!2\left(-E_L,-E_K\right)
\end{equation}
where $\tr\{ \bE\} = 2E_I$ and $\det\{ \bE\} = E_I^2-E_K^2-E_L^2$ are the matrix trace and determinant of $\bE$, and $ \arctan\!2(y,x)$ is the four-quadrant inverse tangent function. Then $\mu$ can be converted back into $\eta$ via the expression in the upper right-hand corner of Table~\ref{eccentricitytable}, if desired. The parameter $\rho^2$ is not particularly interesting here, as it can be absorbed into the choice of constant $c$ in $\bx^T\bE\bx=c^2$.



As the stream function matrix $\bPsi$ for a non-divergent flow and the kinetic energy matrix $\bU^T\bU$ each lacks a $\bJ$ component, like $\bE$ in Equation~(\ref{Eexpansion}), each will describe a family of ellipses, provided, also like $\bE$, that the determinant is nonnegative; otherwise, a family of hyperbolas will be described. Choosing $\bE=\sgn(\zeta)\bPsi$ where $\sgn(\cdot)$ is the signum function, one finds $\tr\{ \bE\} = \frac{1}{2}|\zeta|$ and $\sqrt{\det\{ \bE\}} = \frac{1}{4}\sqrt{\zeta^2-\gamma^2}$ from Equation~(\ref{Psidef}). Contours of the constant stream function are ellipses if $|\zeta|\ge\gamma$, leading to
\begin{equation}
\mu = \frac{|\zeta|}{\sqrt{\zeta^2-\gamma^2}},\quad\quad \eta = \sqrt{\frac{|\zeta|+\gamma}{|\zeta|-\gamma}},\quad\quad\theta = \alpha + \sgn(\zeta)\frac{\pi}{4}
\end{equation}
for the ellipse parameters. The stream function ellipse is therefore circular when the strain vanishes, and~otherwise is oriented with its major axis forty-five degrees from the extensional strain axis. The~ellipse orientation is rotated counterclockwise (in the mathematically-positive sense) from the strain axis if the vorticity $\zeta$ is positive and clockwise (in the mathematically-negative sense) if $\zeta$ is~negative.

For the constant kinetic energy ellipses, we choose $\bE=\bU^T\bU$ and note from Equation~(\ref{kematrix}) that the trace and determinant of $\bU^T\bU$ are given by
\begin{equation}\label{UUtrdet}
\tr\{\bU^T\bU\}=\frac{1}{2}\left(\delta^2+\zeta^2+\gamma^2\right),\quad\quad\sqrt{\det\{\bU^T\bU\}}= \frac{1}{4}\left(\delta^2+\zeta^2-\gamma^2\right).
\end{equation}
The determinant of $\bW\equiv\bU^T\bU $ is readily found if we write $\bU= U_I\bI+ U_J\bJ+ U_K\bK + U_L\bL$, leading to
\begin{multline}\det\left\{\bW\right\} = W_I^2-W_K^2-W_L^2= \left( U_I^2+ U_J^2+ U_K^2+ U_L^2 \right)^2 -4 \left( U_IU_K+ U_JU_L\right)^2 -4 \left( U_IU_L- U_JU_K\right)^2 \\= \left( U_I^2+ U_J^2- U_K^2- U_L^2 \right)^2.
\end{multline}
The determinant expression in Equation~(\ref{UUtrdet}) shows that contours of constant kinetic energy are ellipses provided $\sqrt{\delta^2+\zeta^2}\ge \gamma$. In this case, the extension, aspect ratio, and orientation of the kinetic energy ellipses are found to be 
\begin{equation}
 \mu = \frac{\delta^2+\zeta^2+\gamma^2}{\delta^2+\zeta^2-\gamma^2},\quad\quad\eta = \frac{\sqrt{\delta^2+\zeta^2}+\gamma}{\sqrt{\delta^2+\zeta^2}-\gamma},\quad\quad\theta = \frac{1}{2} \arctan\!2\left(-\delta \gs+\zeta\gn,-\delta \gn-\zeta\gs\right).
\end{equation}
In general, the ellipse orientation depends on the divergence, vorticity, and both components of the strain. However, if the divergence vanishes, the last equation simplifies to 
\begin{equation}
\mu = \frac{\zeta^2+\gamma^2}{\zeta^2-\gamma^2},\quad\quad\eta = \frac{|\zeta|+\gamma}{|\zeta|-\gamma},\quad\quad\theta = \alpha +\sgn(\zeta) \frac{\pi}{4}
\end{equation}
and thus, for non-divergent flow, the kinetic energy ellipses have the same orientation as the stream function ellipses, but a different aspect ratio. The aspect ratio of the kinetic energy ellipses is the square of that stream function's ellipses, implying that the former are more eccentric than the latter. 


\section{Ellipse Kinematics}\label{kinematics}

This section introduces a kinematic model for a particle in a fluid ellipse, and uses it to find the relationships between the rates of change of the ellipse parameters and the spatial derivatives of the linear flow field. A comparison with the kinematic boundary condition approach is also~presented.

\subsection{A Kinematic Model for Fluid Particles in an Ellipse}\label{ellipsemodel}

\begin{figure}[t]
\begin{center}
\includegraphics[width=5.25in,angle=0]{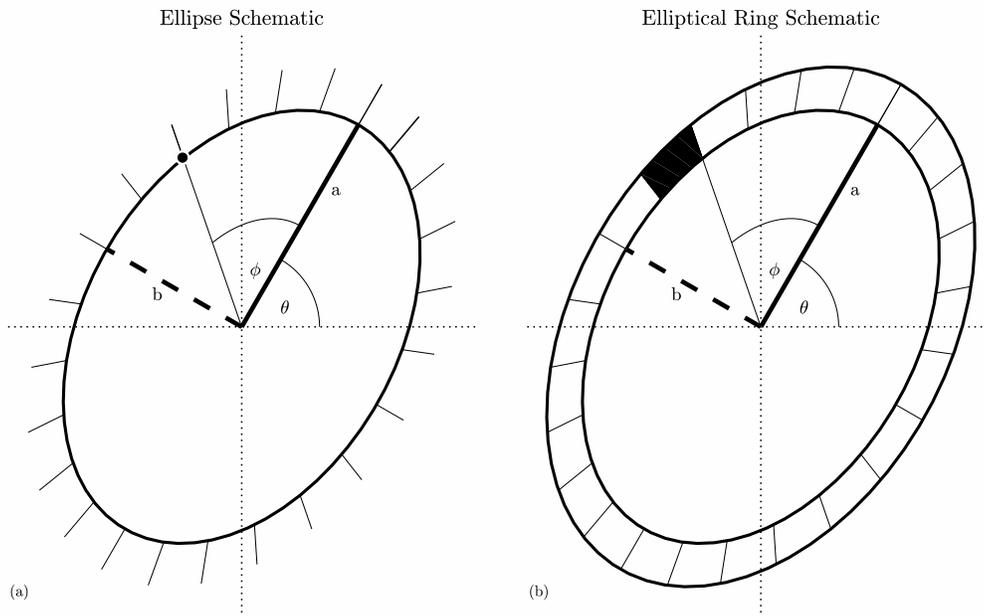}
\end{center}\vspace{-6pt}

	\caption{Schematics for (\textbf{a}) an ellipse and (\textbf{b}) an elliptical ring or annulus. In both panels, the~semi-major and semi-minor axes $a$ and $b$ are denoted by heavy solid and heavy dashed lines, respectively, while the location of one particular particle along the periphery is marked with a thin solid line. The ellipse orientation angle $\theta$, measured counterclockwise from the $x$-axis, and the phase angle $\phi$ to the particle location from the major axis are also shown. The twenty-four ``spokes'' around the ellipse mark uniform increments of $\Delta\phi=\pi/48$ radians in the phase angle $\phi$. In ({b}), two~concentric ellipses of identical shape and orientation, but slightly different sizes, are drawn. The space between these two ellipses forms an \emph{elliptical ring} or \emph{annulus}, with the phase angle $\phi$ now taken to mark \emph{parcels}, shown later to have the same area, as opposed to simply a location around the ellipse periphery. The~parcel between the indicated phase angle $\phi$ and $\phi+\Delta\phi$ is filled in with dark shading.}\label{schematic}
\end{figure}

A parametric representation of the time-varying position of a fluid particle located on an ellipse with semi-major axis $a(t)$, semi-minor axis $b(t)\le a(t)$, and major axis orientation $\theta(t)$ is given by 
\begin{equation}\label{kinematicmodel}
\breve\bx(t)\equiv\breve\bx\left(a(t),b(t),\theta(t),\phi(t)\right) \equiv \bR\left(\theta\right) \begin{bmatrix}a \cos\phi\\b\sin\phi\end{bmatrix}
\end{equation}
where the phase $\phi(t)$ controls the position of the particle around the ellipse periphery. The breve symbol will be used to indicate that $\breve\bx(t)$ is a Lagrangian quantity, as distinct from the Eulerian position vector $\bx$. A schematic is shown in Figure~\ref{schematic}a. This parametric model can either be thought of as representing a \emph{particle}, by which we mean an infinitesimal point along an ellipse, or~else a deformable fluid \emph{parcel} of small but finite volume between two concentric ellipses, as shown in Figure~\ref{schematic}b. This subtle distinction is not important in this section, but will be important in the following section when the physical properties of the fluid ellipse are computed. It is important to emphasize that Equation~(\ref{kinematicmodel}) does not specifically represent some special set of points, such as a vortex boundary; rather, it can denote any set of particles that one may wish to mark out as an ellipse within a linear flow field.

A comment should also be made regarding the phase angle $\phi$. This not the same as the geometric azimuth or polar angle $\Theta\equiv\arctan(\breve y/\breve x)$, where $\breve x$ and $\breve y$ are the components of $\breve \bx(t)$. The azimuth angle, measured in a rotated reference frame in which the ellipse orientation angle $\theta$ is zero, becomes $\widetilde\Theta =\arctan\left((b/a) \tan\phi\right)$, which is clearly different from $\phi$ except for the circular case in which $a$ and $b$ are equal. In Figure~\ref{schematic}a, the angle subtended by $\phi$ is marked with an elliptical arc, rather than a circular arc, in order to emphasize this distinction; this corrects a shortcoming of the ellipse schematics of \citet{lilly06-npg} and \citet{lilly10-itsp}, pointed out to the author by S. Elipot.

The velocity of the particle, or parcel, described by the vector $\breve\bx(t)$ will be denoted $\breve\bu(t)\equiv \frac{\rd}{\rd t}\breve\bx(t)$, again as distinct from the Eulerian velocity due to a linear velocity field, $\bu(\bx,t)=\bU(t)\bx$. Here it will 
be convenient to reparametrize the Lagrangian ellipse vector $\breve\bx(t)$ in terms of the geometric mean radius $\rho=\sqrt{a b}$ and aspect ratio $\eta=a/b$. Then $a=\rho\sqrt{\eta}$ and $b=\rho/\sqrt{\eta}$, leading to
\begin{equation}
\breve\bx(t)=\breve\bx(\rho(t),\eta(t),\theta(t),\phi(t)) 
= \frac{\rho}{\sqrt{\eta}}\,\bR\left(\theta\right) \begin{bmatrix}\eta\cos\phi\\ \sin\phi \end{bmatrix}.\label{kinematicmodeleta}
\end{equation}
Taking the time derivative of this expression, the particle velocity $\breve \bu(t)$ is readily found to be
\begin{equation}\label{rotatedvector}
\breve\bu(t) =
\frac{\rho}{\sqrt{\eta}}\bR(\theta)\left\{\frac{\rd \ln \rho}{\rd t} \begin{bmatrix} \eta\cos\phi\\ \sin\phi \end{bmatrix} +\frac{1}{2}\frac{\rd\ln \eta}{\rd t} \begin{bmatrix} \eta\cos\phi\\ -\sin\phi \end{bmatrix}+\frac{\rd\theta}{\rd t}\begin{bmatrix}-\sin\phi \\\eta \cos\phi\end{bmatrix}
+\frac{\rd\phi}{\rd t}\begin{bmatrix}-\eta \sin\phi\\\cos\phi\end{bmatrix}\right\}
\end{equation}
where $\frac{\rd}{\rd t}\ln\rho(t)$ will be called the \emph{expansion rate}, the rate of change of the aspect ratio $\frac{\rd}{\rd t}\ln\eta(t)$ is one of several possible measures of the \emph{deformation rate}, $\frac{\rd}{\rd t}\theta(t)$ is the \emph{precession rate}, and $\frac{\rd}{\rd t}\phi(t)$ is referred to as the \emph{orbital frequency}. 

\subsection{The Ellipse Flow Matrix}

The velocity $\breve\bu(t)$ of a particle along the ellipse periphery may be equivalently expressed through the matrix multiplication $\breve\bu(t)= \breve\bU(t) \breve\bx(t)$, where the \emph{ellipse flow matrix} $\breve \bU(t)$ is defined by
\begin{equation}\label{ellipseflowmatrix}
\breve\bU(t)\equiv\bR(\theta)\left\{
\frac{\rd \ln \rho}{\rd t}\bI +\left(\frac{\rd\theta}{\rd t}+\frac{\eta^2+1}{2\eta}\frac{\rd\phi}{\rd t}\right)\bJ+ \frac{1}{2}\frac{\rd\ln \eta}{\rd t}\bK-\frac{\eta^2-1}{2\eta}\frac{\rd\phi}{\rd t}\bL\right\}\bR^T(\theta).
\end{equation}
Whereas $\breve\bu(t)$ only gives the velocity of one particle along the ellipse, the matrix $\breve\bU(t)$ describes the entire linear flow field implied by the evolution of the parametric ellipse model at any position $\bx$. The~ellipse flow matrix may be rewritten in the somewhat more transparent form
\begin{equation}\label{ellipseflowmatrixsimple}
\breve\bU(t)\equiv
\bR(\theta)\left\{\frac{\rd \ln \rho}{\rd t}\bI + \frac{\rd\theta}{\rd t}\bJ+ \frac{1}{2}\frac{\rd\ln \eta}{\rd t} \bK
+\frac{\rd\phi}{\rd t}\begin{bmatrix} 0 & -\eta\\ 1/\eta & 0\end{bmatrix}\right\}\bR^T(\theta)
\end{equation}
from which one sees at once that $\breve\bU(t)\breve \bx(t)$, with $\breve \bx(t)$ defined by Equation~(\ref{kinematicmodeleta}), does indeed recover the expression for $\breve \bu(t)$ given by Equation (\ref{rotatedvector}).  In  deriving the latter equation from Equation~(\ref{ellipseflowmatrix}), it is helpful to note the identity
$\frac{\eta^2+1}{2\eta}\bJ-\frac{\eta^2-1}{2\eta}\bL=\begin{bmatrix} 0 & -\eta \\ 1/\eta & 0\end{bmatrix}$.

Equation~(\ref{ellipseflowmatrixsimple}) for $\breve \bU(t)$ has a clear physical interpretation. The velocity implied by the first term, involving the expansion rate $\frac{\rd}{\rd t}\ln\rho(t)$, changes the magnitude of the ellipse vector $\breve\bx(t)$ without changing its angle. The velocity implied by the second term, involving the precession rate $\frac{\rd}{\rd t}\theta(t)$, is always oriented perpendicular to $\breve\bx(t)$, expressing a tendency for solid-body rotation. The velocity implied by the third term, involving the deformation rate $\frac{\rd}{\rd t}\ln\eta(t)$, acts for the positive deformation rate to increase the major axis $a$ while at the same time decreasing the minor axis $b$. The final matrix term, multiplying the orbital frequency $\frac{\rd}{\rd t}\phi(t)$, transforms $\left[\eta\cos\phi \,\, \sin \phi\right]^T$ into $\left[-\eta\sin\phi \,\,\cos \phi\right]^T$, i.e., it advances the phase $\phi$ by ninety degrees; this implies a velocity that is tangent to the ellipse periphery, as will be seen more clearly later; see Equation~(\ref{elldef}) of Section~\ref{integrals}.


\subsection{The Ellipse Evolution Equations and Ellipse/Flow Equivalence}

The rates of change of the parameters of the ellipse model can be directly linked to the properties of a linear flow field, leading to explicit expressions for those rates of change. With $\tilde\alpha\equiv \alpha-\theta$ being the orientation of the strain axis in the reference frame of the ellipse, one finds at once 
\begin{equation}\label{correspond}
\delta=2\frac{\rd \ln \rho}{\rd t},\quad\quad \zeta=2\left(\frac{\rd\theta}{\rd t}+\frac{\eta^2+1}{2\eta}\frac{\rd\phi}{\rd t}\right),\quad\quad \gamma\cos2\tilde\alpha =\frac{\rd\ln \eta}{\rd t},\quad\quad \gamma\sin2\tilde\alpha = -2\frac{\eta^2-1}{2\eta}\frac{\rd\phi}{\rd t}
\end{equation}
by comparing Equation~(\ref{ellipseflowmatrix}) for the flow matrix $\breve\bU(t)$ implied by the Lagrangian velocity $\breve\bu(t)$, with~Equation~(\ref{Udecomp2}) for a general flow matrix $\bU(t)$. These can be readily combined to give expressions for the strain magnitude $\gamma$ and angle $\alpha$, if desired, thus establishing a correspondence between the velocity gradient quantities $\delta$, $\zeta$, $\gamma$, and $\alpha$ of the linear velocity field, and the rates of change of the ellipse parameters. Rearranging Equation~(\ref{correspond}) leads to
the \emph{ellipse evolution equations}, given by
\begin{equation}\label{etol}
\frac{\rd \ln \rho}{\rd t}= \frac{1}{2}\delta,\quad \quad 
\frac{\rd\ln \eta}{\rd t} = \gamma \cos2\tilde\alpha,\quad \quad 
\frac{\rd\theta}{\rd t}= \frac{1}{2}\zeta +\frac{1}{2}\frac{\eta^2+1}{\eta^2-1} \gamma \sin2\tilde\alpha,\quad\quad 
\frac{\rd \phi}{\rd t}= -\frac{1}{2}\frac{2\eta}{\eta^2-1} \gamma \sin2\tilde\alpha
\end{equation}
which determine the evolution of the parameters of any ellipse in a possibly time-dependent linear flow. Observe that the left-hand sides are all Lagrangian rates of change, while the right-hand sides contain a mixture of Eulerian properties of the flow field and Lagrangian properties of the ellipse. 

The evolution equations state that the divergence sets the fractional rate of change of the geometric mean radius, $\frac{\rd}{\rd t}\ln\rho(t)$; the normal strain in the reference frame of the ellipse controls the deformation $\frac{\rd}{\rd t}\ln \eta(t)$; the vorticity contributes to the precession rate $\frac{\rd}{\rd t}\theta(t)$; and the shear strain in the reference frame of the ellipse contributes to the precession rate and also entirely controls the orbital frequency $\frac{\rd}{\rd t}\phi(t)$. Perhaps surprisingly, the vorticity does not contribute to the shifting of the position of particles around the ellipse via the orbital frequency. Instead, it is the shear strain in the ellipse reference frame that accomplishes this. Examining the sign of the strain angle relative to the ellipse orientation, $\tilde\alpha=\alpha-\theta$, we see that strain always acts to rotate the vortex toward the extending axis of the strain; recall here that $\gamma$ is the strain magnitude and is therefore non-negative. This is accompanied by a shift in phase angle $\phi$ that causes the particles to circulate along the ellipse in the opposite direction from the strain-induced rotation of the ellipse itself. 

Equations~(\ref{correspond}) and (\ref{etol}) together mean that given the evolution of any nondegenerate ellipse, we know the instantaneous properties of the unique linear flow that could generate such evolution, and conversely, given the properties of a linear flow, we know how any ellipse advected by that flow will evolve. (By ``nondegenerate'' we mean to exclude the singular cases of a line, for which $\eta=\infty$, and a circle, for which $\eta=1$, as in both of these cases, the ellipse model only has three free parameters instead of four.) This result is referred to as the principle of \emph{ellipse/flow equivalence}. It~reflects a deep correspondence between Eulerian and Lagrangian quantities for the case of a linear flow. The~instantaneous properties of an arbitrary linear flow field and the instantaneous rates of change of an elliptical ring of fluid, therefore, contain equivalent information.

Two caveats should be mentioned at this point. The first is that ellipse/flow equivalence does not mean that if a fluid ellipse evolves at one particular moment into a new ellipse, the flow that caused this is necessarily linear; it means that \emph{if} we know that the flow is linear, then the ellipse evolution also tells us the flow properties. This subtle point will be returned to later. Secondly, we have not specified how the parameters of the ellipse model and their rates of change are to be inferred. An interesting question, relevant to the Lagrangian analysis methods of e.g. \cite{lilly11-grl}, is whether the ellipse parameters can be accurately estimated from a single Lagrangian trajectory. Such issues are outside the scope of the present paper, where we simply ask what we can do with the knowledge of the rates of change of the ellipse parameters, if this information is available. 


The ellipse evolution equations are not themselves new. They are essentially those of the Kida vortex \cite{kida81-jpsj,neu84-pf,ide95-fdr} if the vorticity anomaly is set to zero, but with the addition of the divergence and the rate of change of phase. The derivation used in those works involves considering the kinematic condition for the advection of an elliptical boundary, as discussed in more detail shortly. That approach does not lead to an equation for the orbital phase $\phi(t)$, though it could be inferred from conservation of circulation. What is new is the understanding of the intimate link between ellipse evolution and linear flows as a \emph{general} result, which has been made apparent here through the use of the four-parameter kinematic ellipse model, as opposed to the three-parameter kinematic boundary condition. 

Note that the ellipse evolution equations do not prohibit the aspect ratio $\eta$ from evolving to be less that one; but we have assumed $a\ge b$ and, therefore, $\eta>1$. This difficulty can be addressed as follows. If $\eta=1$, the ellipse becomes a circle, and the orientation $\theta$ is thus undefined. At moments for which $\eta=1$, we are free to introduce discontinuities of $\pi/2$ in the orientation angle $\theta$ such that subsequent evolution will continue with $\eta\ge1$. That is, at any times at which the ellipse momentarily becomes a circle, we may relabel the axes in order that $a$ always refers to the longer of the two axes. It~turns out that this approach is sufficient for most cases, e.g., the Kida vortex solution, which does not cross $\eta=1$ in finite time. For cases in which this singularity is problematic, it may be removed by a~modified choice of elliptical parameters; see the note on p. 852 of \cite{legras91-pfa}.

\subsection{The Kinematic Boundary Condition Approach}\label{kinematicsection}


Here, the ellipse evolution equations have been derived with the use of a parametric model of a particle orbiting a time-varying ellipse. A more standard approach in the literature, that used by, e.g., \citet{kida81-jpsj}, \citet{neu84-pf}, and \citet{ide95-fdr}, is to consider the kinematic boundary condition for an evolving elliptical curve; this approach that may be extended to handle ellipsoids in three dimensions, see \citet{mckiver03-jfm}. The kinematic condition for an evolving elliptical boundary is described here for comparison with the parametric method.

With $\bE(t)$ being the ellipse matrix given in Equation~(\ref{ellipsematrix}), we define $\chi(\bx,t)\equiv\bx^T\bE(t)\bx$, with~contours of constant $\chi$ being ellipses. In order for a particle to remain on the same ellipse, as it is advected following the flow, it must be the case that the total time derivative of $\chi$~vanishes: 
\begin{equation}\label{kinematic}
\frac{D}{Dt}\chi(\bx,t)=\left(\frac{\partial }{\partial t} + \bu \cdot \nabla\right) \chi=\left(\frac{\partial }{\partial t} + \bu^T \bm{\nabla}\right) \left(\bx^T \bE \bx\right) = 0.
\end{equation}
Following \cite{neu84-pf}, this may be rearranged to give
\begin{equation}\label{kinematicsymmetric}
\bx^T\left(\frac{\rd \bE}{\rd t} + \bU^T\bE+\bE^T\bU \right) \bx = 0
\end{equation}
where we have observed from inspection that $\bm\nabla \left(\bx^T \bG \bx\right)=\left(\bG+\bG^T\right)\bx$ for some spatially-constant real-valued matrix $\bG$. Note that it is important to arrange the quantity in parenthesis in Equation~(\ref{kinematicsymmetric}) such that it is symmetric. The solution is then found by choosing
\begin{equation}\label{matrixequation}
\frac{\rd \bE}{\rd t} = -\bE\bU - \bU^T \bE 
\end{equation}
where we have noted $\bE=\bE^T$. From this, one may deduce the first three evolution equations, as~shown the Appendix \ref{kinematicappendix}. This is done in \citet{kida81-jpsj}, \citet{neu84-pf}, and \citet{ide95-fdr} for the case of the Kida~vortex, whereas here we consider an arbitrary linear flow field. As pointed out above, an~equation for the rate of change of the phase $\phi$ is not obtained, as the kinematic boundary condition does not track the locations of particles around the ellipse periphery.


\section{Integrals of a Fluid Ellipse}\label{integrals}

In this section, expressions for various integral properties of a fluid ellipse are derived. The~relationship between integrals over a thin elliptical \emph{ring} or \emph{annulus} of fluid, versus those over an~elliptical \emph{disk} of fluid, is discussed. The kinetic energy averaged around the elliptical ring is shown to have an interesting partitioning into three distinct physical terms, an apparently new result.


\subsection{An Elliptical Ring of Fluid}


Rather than merely describing a position along the periphery of any ellipse, as in Figure~\ref{schematic}a, we~let $\breve\bx(t)$ describe the location of a fluid parcel within a thin elliptical ring or annulus, as in Figure~\ref{schematic}b, around which parcels may flow while preserving their volume.  (The terms ``ring'' and ``annulus'' will be used interchangeably herein, as ``annulus'' means ``ring'' in Latin.) The ring could be imagined as a~deformable tank of negligible weight. It is taken to have a uniform (but possibly time-variable) height $h(t)$ measured in the dimension out of the page, and a width that is proportional to the distance from the origin, i.e., a width of $\varepsilon \|\breve \bx(t)\|$ for some small positive number $\varepsilon(t)$. The fluid is assumed to have a constant density $\varrho$. The ring volume $V_R$ is therefore constant and is given by $V_R=A_R h$, where $A_R=2\pi a b\varepsilon$ is the ring area.  Conservation of volume implies that the ring width $\varepsilon$ must be equal to $\varepsilon = V_R/(2\pi a b h)$. The annulus may change orientation, shape, or size, consistent with volume preservation. The reason for permitting a time-varying height is so that the integrals will be applicable both to the two-dimensional Euler and shallow water systems. 

An important point is that the elliptical annulus is different from, for example, an elliptical wire, which could also be described by $\breve\bx(t)$, but which will have a different mass density and therefore different integrals. Computing, for example, the moment of inertia of an elliptical wire, one would need to evaluate elliptic integrals (also known as elliptic functions), which will not be needed here. 

The ellipse integrals will be derived for a non-rotating reference frame, but can trivially be modified for a rotating reference frame. Because all of the properties to be derived are kinematic (that is, not referring to any forces), there is no difference between a rotation of the reference frame and a rotation of the ellipse. In order to be applicable to an ellipse lying within a reference frame that is rotating about the vertical axis at a rate of $\frac{1}{2}f$, where $f$ is the Coriolis frequency, one simply formally replaces the ellipse rotation rate $\frac{\rd}{\rd t}\theta(t)$ with $\frac{\rd}{\rd t}\theta(t)+\frac{1}{2}f$ wherever it appears in the following expressions. Then the angular momentum, circulation, and kinetic energy derived below become absolute quantities as measured in the non-rotating frame. 

Because the fluid density $\varrho$ is constant and the thickness $h$ is assumed to be spatially uniform, one~finds with $\rd A$ being a differential area 
\begin{equation}
\frac{1}{\varrho V_R} \varrho h \,\rd A = \frac{1}{V_R} \,h\,\rd A = \frac{1}{A_R} \, \rd A
\end{equation}
and mass-weighted area averages are therefore the same as simple area averages. For simplicity in what follows, we will write averaging integrals in the latter form.

\subsection{Moment of Inertia, Angular Momentum, and Circulation}

Three important physical properties of an evolving elliptical ring of fluid, occupying an annular region $R$ (for ``ring'') in a flat domain and bounded on the exterior by a curve $C$, are defined as 
\begin{equation}\label{threeintegrals}
I(t) \equiv \frac{1}{A_R}\iint_R \|\bx\|^2 \,\rd A,\quad\quad
M(t) \equiv \frac{1}{A_R} \iint_R \,\bk \cdot \left(\bx\times\bu\right)\, \rd A,\quad\quad
\Pi(t) \equiv 
 \frac{1}{2\pi}\oint_{C} \bu\cdot\rd\bx 
\end{equation}
which are, respectively, the moment of inertia per unit mass, the average angular momentum per unit mass, and $1/(2\pi)$ times the enclosed circulation. Here the contour integral in the final equation is taken in the right-hand sense around the contour $C$, with $\rd\bx$ being a differential segment of $C$. The~quantity $\Pi(t)$ is related to the usual circulation $\Gamma(t)$ by $\Pi\equiv \Gamma/(2\pi)$, and is therefore recognized as the angular density of the circulation. We will work with $\Pi$ rather than $\Gamma$ in order to emphasize a~similarity to the angular momentum. These three integrals become
\begin{equation}\label{threeintegralvalues}
I(t) = \frac{a^2+b^2}{2},\quad\quad
M(t) =\frac{a^2+b^2}{2} \frac{\rd\theta}{\rd t} + ab\frac{\rd\phi}{\rd t},\quad\quad
\Pi(t) 
 =ab\frac{\rd\theta}{\rd t} + \frac{a^2+b^2}{2}\frac{\rd\phi}{\rd t}
\end{equation}
for the evolving elliptical ring, as will be shown shortly. Note that deformation and change in the ellipse area contribute neither to the angular momentum, nor to the circulation. 

Observe the symmetry in form between the ellipse angular momentum $M$ and the normalized circulation $\Pi$. These expressions for $M$ and $\Pi$ are essentially the same as (4.6b) of \citet{holm91-jfm}, in which these quantities emerge from a Hamiltonian framework as the canonical momenta conjugate to the orientation angle $\theta$ and orbital phase $\phi$, respectively, for a shallow-water elliptical vortex. $M$ and $\Pi$ are two distinct but related quantities, both with units of length squared per unit time, or angular momentum per unit mass. These both describe the rotation of the system, but in two different ways.

To clarify the distinction between $M$ and $\Pi$, we first consider the case of solid-body rotation of the ellipse, for which $\frac{\rd}{\rd t}\phi(t)$ vanishes. Let $\Omega\equiv\frac{\rd}{\rd t}\theta(t)$ be the ellipse rotation rate. Then the angular momentum per unit mass is enhanced over the angular rotation rate through the moment of inertia, $M=I \Omega$, as usual. A solid-body velocity given by $ \bu=\Omega\bk\times\bx=\Omega\bJ\bx=\Omega\,[- y \,\,\, x]^T$ corresponds to a vorticity of $\zeta=\frac{\rd }{\rd x}v-\frac{\rd }{\rd y}u= 2\Omega$. From Stokes' theorem, the circulation is the spatially-integrated vorticity within the enclosed contour, so we have $\Gamma=2\Omega \pi ab$ for solid-body rotation, or $\Pi= \Omega \,a b$, in agreement with Equation~(\ref{threeintegralvalues}). Thus the first term in both $M$ and $\Pi$ is due to the effect of solid-body rotation. The second term similarly captures the angular momentum $ab\frac{\rd}{\rd t}\phi(t)$ and circulation $I\frac{\rd}{\rd t}\phi(t)$ associated with the flow of particles along the elliptical ring. Note that with the ellipse area fixed, the~term proportional to $I$ can increase without bound through an elongation of the ellipse, increasing the $\theta$ contribution to $M$ and the $\phi$ contribution to $\Pi$. More generally, conservation of $M$ or of $\Pi$ would imply relationships between the ellipse geometry and the rates of change of these two angles. 

The above expressions for $M$ and $\Pi$ are general, in the sense that they are expressed in terms of the rates of change of the ellipse itself. For the particular case of an ellipse in a linear flow, one may substitute from the ellipse evolution equations, Equation~(\ref{etol}), to find
\begin{equation}\label{MPvaluesflow}
M(t) =\frac{1}{2}\rho^2\left[\frac{\eta^2+1}{2\eta}\zeta+\frac{\eta^2-1}{2\eta} \gamma \sin2\tilde\alpha\right],\quad\quad
\Pi(t) 
=\frac{1}{2}\rho^2\zeta.
\end{equation}
Differentiating these expressions and substituting again from Equation~(\ref{etol}) leads to 
\begin{equation}
\frac{\rd M}{\rd t} = M \frac{\rd \ln \rho^2}{\rd t} +\frac{1}{2} \rho^2 \frac{\eta^2-1}{2\eta}\left(\frac{\rd \gamma}{\rd t} \sin2\widetilde \alpha - 2\gamma \frac{\rd \alpha}{\rd t} \cos2\widetilde \alpha\right)
 \label{dMdt},\quad\quad
 \frac{\rd \ln \Pi}{\rd t} = \frac{\rd \ln \rho^2}{\rd t} + \frac{\rd \ln \zeta}{\rd t}
\end{equation}
after making use of the derivative relationships $\frac{\rd}{\rd t}\left[(\eta^2\pm1)/2\eta\right]=\left[(\eta^2\mp 1)/2\eta\right] \frac{\rd}{\rd t}\ln \eta$. Thus, in a~non-divergent linear flow that is constant in time, and in the absence of any vorticity anomalies, both the average angular momentum as well as the circulation of any fluid ellipse are constant. While the latter conservation law is well known, the former is perhaps surprising. The term proportional to $\frac{\rd}{\rd t}\ln \rho^2$ in $\frac{\rd }{\rd t}M$ deserves comment, as one would expect angular momentum to be conserved independent of changes in the size of the ring. It is important to keep in mind that the ambient flow is considered to be imposed. Thus, for example, if the strain vanishes, but the vorticity is held fixed, then $\frac{\rd }{\rd t} \ln M= \frac{\rd}{\rd t} \ln \rho^2$ gives the correct conservation law; the angular momentum increases as the ring area increases on account of the fixed vorticity. 

For a non-divergent flow, another important physical property is the spatially-averaged value of the stream function over the elliptical ring, which is found to be given by
\begin{equation}\label{averagepsi}
\overline \psi(t) \equiv \frac{1}{A_R} \iint_R \psi \, \rd A = \frac{1}{2}\left\{\frac{a^2+b^2}{2} \frac{\rd\theta}{\rd t}+ab\frac{\rd\phi}{\rd t}\right\} +\psi_o= \frac{1}{2}M+\psi_o.
\end{equation}
This equals one half of the angular momentum per unit mass of the elliptical ring, plus the spatially-uniform portion of the stream function, $\psi_o(t)$, as follows directly from the pointwise relationship between the stream function and angular velocity described earlier in Equation~(\ref{psiandangular}).

\subsection{Physical Properties of an Elliptical Disk of Fluid}\label{diskintegrals}

The integrals of an elliptical \emph{disk} bear a simple relation to the integrals of an elliptical \emph{ring}, which~we assume to have a spatially uniform but potentially time-varying height $h$. For a quadratic quantity, given by $\bx^T\bG\bx$ for some matrix $\bG$, it will be shown later that (with ``$D$'' denoting the disk)
\begin{equation}
\frac{1}{A_D}\iint_D \,\bx^T \bG \bx \,\rd A = \frac{1}{2}\left(\frac{1}{A_R}\iint_{R} \,\bx^T \bG \bx \,\rd A \right)
\end{equation}
such that the mass-weighted average value of $\bx^T\bG\bx$, taken over the entire elliptical disk, is simply one half its mass-weighted average value over one of the concentric elliptical rings $R$ within the disk $D$. In other words, the ``disk integrals'' are one half of the ``ring integrals'' for quadratic quantities, which includes $I$ and $M$, as well as the kinetic energy $K$ introduced in the next section. 

The stream function is an exception. Because it includes a quadratic term, as well as a spatially-uniform term, the disk-averaged stream function is no longer equal to one half of the ring-averaged stream function. Instead, one finds
\begin{equation}
\frac{1}{A_D}\iint_D \, \psi(\bx,t) \,\rd A 
=\frac{1}{A_D}\iint_{D} \,\, \bx^T \bPsi \bx \,\rd A + \psi_o = \frac{1}{2}\left(\frac{1}{A_R}\iint_{R} \,\psi(\bx,t) \,\rd A\right) +\frac{1}{2}\psi_o
\end{equation}
where we note the appearance of an extra additive constant with a value of $\frac{1}{2}\psi_o$ in the disk-averaged version compared with the ring average. This distinction will be important for interpreting the so-called ``excess'' domain-integrated kinetic energy of an elliptical two-dimensional vortex such as the Kida vortex, which is based on the value of the stream function; see e.g. \cite{vanneste10-pf}.

Note that these results on disk averages directly apply to vortices in two-dimensional flow, such~as the Kida vortex, but not to the shallow water vortex solutions, in which the ellipse height exhibits a~quadratic dependence on the spatial coordinates.



\subsection{Kinetic Energy}

Along with the integrals presented in the last section, we may also compute the average kinetic energy per unit mass experienced by parcels along the elliptical ring, defined as
\begin{equation}\label{keintegral}
K(t) \equiv \frac{1}{A_R} \iint_R \frac{1}{2} \|\bu\|^2\, \rd A.
\end{equation}
In terms of the geometric mean radius $\rho=\sqrt{ab}$, aspect ratio $\eta=a/b$, orientation $\theta$, and phase $\phi$, this will be found to be 
\begin{multline} \label{kexpression}
K(t)= \frac{1}{2}\rho^2\frac{\eta^2+1}{2\eta}\left\{ \left(\frac{\rd\phi}{\rd t}+\frac{2\eta}{\eta^2+1}\frac{\rd\theta}{\rd t}\right)^2+\left(\frac{\rd \ln \rho}{\rd t}+\frac{1}{2}\frac{\eta^2-1}{\eta^2+1}\frac{\rd\ln\eta}{\rd t}\right)^2\right.\\\left.
+\left(\frac{\eta^2-1}{\eta^2+1}\right)^2\left(\frac{\rd\theta}{\rd t}\right)^2+\left(\frac{2\eta}{\eta^2+1}\right)^2\left(\frac{1}{2}\frac{\rd\ln \eta}{\rd t}\right)^2\right\}
\end{multline}
an expression that may be simplified through the choice of different variables. Observe that the fractional rate of change of the ring's moment of inertia, $I$, may be expanded to give 
\begin{equation} \label{kexpressionI}
\frac{1}{2}\frac{\rd \ln I}{\rd t} =
\frac{1}{2} \frac{\rd}{\rd t} \ln\left(\rho^2\frac{\eta^2+1}{2\eta}\right)
=\frac{\rd \ln \rho }{\rd t} + \frac{1}{2}\frac{\eta^2-1}{\eta^2+1}\frac{\rd \ln\eta}{\rd t}
\end{equation}
which is the second term in parenthesis in Equation~(\ref{kexpression}). Replacing $\rho$ with the moment of inertia $I$ and $\eta$ with the linearity $\lambda = (\eta^2-1)/(\eta^2+1)$, one finds
\begin{equation} \label{kefourterms}
K(t)= \frac{1}{2}I\left\{ \left(\frac{\rd\phi}{\rd t}+\sqrt{1-\lambda^2}\,\frac{\rd\theta}{\rd t}\right)^2+\left(\frac{1}{2}\frac{\rd \ln I}{\rd t}\right)^2
+\lambda^2\left(\frac{\rd\theta}{\rd t}\right)^2+\frac{1}{1-\lambda^2}\left(\frac{1}{2}\frac{\rd\lambda}{\rd t}\right)^2\right\}
\end{equation}
after making use of Table~\ref{eccentricitytable} together with the derivative relations in Equation~(\ref{derivrelations}). 

This expression has a very interesting interpretation. The entire first term is found to be simply the squared normalized circulation $\Pi$ divided by the moment of inertia $I$:
\begin{equation}
 \frac{1}{2}I \left(\frac{\rd\phi}{\rd t}+\sqrt{1-\lambda^2}\,\frac{\rd\theta}{\rd t}\right)^2 = \frac{1}{2}\frac{\Pi^2}{I}.
\end{equation}
If circulation is conserved, this term reflects the generation of kinetic energy through changing the moment of inertia, that is, the average squared distance of fluid parcels from the origin. This process could be called \emph{kinetic energy induction}. The second term in Equation~(\ref{kefourterms}) is due to the deformation and expansion/contraction velocities that contribute to changing the moment of inertia. The third term is the additional contribution of precessional velocities not already included in the induction term, and the fourth term is the additional contribution of deformational velocities not already included in the moment of inertia term. Later, the sum of the third and fourth terms will be shown to have an interesting interpretation: it represents the ring variance of parcel angular velocity, that is, the average squared departure of the angular velocity from its average value along the ring. 



The expression for the average kinetic energy of an elliptical ring in Equation~(\ref{kexpressionI}) can be rearranged to give
\begin{equation} \label{kefourtermsball}
 \frac{1}{2}\left\{ \left(\frac{\rd I}{\rd t}\right)^2+
I^2\left[4\lambda^2\left(\frac{\rd\theta}{\rd t}\right)^2+\frac{1}{1-\lambda^2}\left(\frac{\rd\lambda}{\rd t}\right)^2\right]\right\} =4KI - 2 \Pi^2 
\end{equation}
a form that is strikingly similar to Equation~(22) of \citet{ball63-jfm} for the evolution of the moment of inertia of a body of fluid lying on a paraboloid, and to Equation~(2.3) of \citet{young86-jfm} for the evolution of the moment of inertia of a shallow-water elliptical vortex. In both of those systems, the entire quantity in square brackets is replaced with a constant, $f^2$, the squared Coriolis frequency, and the moment of inertia is found to oscillate at that frequency. Here, one observes that \emph{if} the orientation angle and linearity evolve such that the term in square brackets is a constant, say $\omega_o^2$, and if the circulation $\Pi$ and kinetic energy $K$ are both also constant, then a decoupling occurs. Under these conditions, one may differentiate Equation~(\ref{kefourterms}) to give
\begin{equation} 
 \frac{\rd^2 I}{\rd t^2}+\omega_o^2I=4K
\end{equation}
and the moment of inertia oscillates as a harmonic oscillator, independent of the other ellipse parameters. This is again reminiscent of the results of \citet{ball63-jfm} and \citet{young86-jfm}, suggesting that the moment of inertia oscillations seen therein are a reflection of a kinematic constraint. Further exploration of the relationship between this kinematic analysis and the moment of inertia oscillations found in those two physical systems is left to the future. 


\subsection{Computing the Integrals of the Ellipse}

To compute the integrals of the elliptical ring, we will need properties all along the ellipse at each time. Therefore we reparametrize the ellipse vector $\breve\bx(t)$ in terms of a free phase $\varphi$ as 
 \begin{equation}\label{xfreephase}
\breve \bx(\varphi,t)\equiv\breve \bx(\rho(t),\eta(t),\theta(t),\varphi)
\end{equation} 
where $\varphi$, which can be chosen to access the location of any parcel along the fluid ellipse, is distinguished from $\phi(t)$, a time-varying property of one particular parcel. Similarly, one may define a version of the velocity $\breve\bu(t)$ that is also dependent on a free phase as
 \begin{equation}\label{ufreephase}
\breve \bu(\varphi,t)\equiv\breve \bU \breve\bx(\varphi,t)
\end{equation} 
which gives the velocity at the location of any parcel along the ellipse through a suitable choice of the phase $\varphi$. In general, $\breve \bx(\varphi,t)$ and $\breve \bu(\varphi,t)$ parameterized by a free phase will only appear inside integrals over $\varphi$; thus, there is no danger of confusing them with $\breve \bx(t)$ and $\breve \bu(t)$.

We will also need a differential vector $\rd\breve\bx(\varphi,t)$ that is tangent to the ellipse periphery. Differentiating the ellipse vector with a free phase, $ \breve\bx(\varphi,t)$, with respect to the phase $\varphi$, we obtain 
\begin{equation}\label{elldef}
\rd\breve\bx(\varphi,t) \equiv \frac{\partial }{\partial \varphi}\breve\bx(a,b,\theta,\varphi)\, \rd\varphi 
=\bR\left(\theta\right) \begin{bmatrix}- a \sin\varphi\\b\cos\varphi \end{bmatrix}\rd\varphi.
\end{equation}
A differential unit of area of the elliptical ring as a function of the free phase $\varphi$ is then given by
\begin{equation}\label{diffarea}
\rd A = \| \varepsilon\breve \bx \times \rd \breve\bx \|= \varepsilon\left\| \breve \bx \times \frac{\partial\breve \bx}{\partial \varphi} \, \rd \varphi\right\| = \varepsilon\left| \breve \bx^T \bJ^T \frac{\partial\breve \bx}{\partial \varphi} \, \rd \varphi\right| = a b \varepsilon \,\rd \varphi
\end{equation}
after making use of Equation~(\ref{crossproduct}) for the cross product of two two-vectors. Here $\varepsilon$, as described earlier, is~a~small number specifying the width of the fluid annulus as $\varepsilon\|\breve \bx\|$, i.e. with a width at each point that is proportional to the distance from the origin. The area of the elliptical ring is found to be 
\begin{equation}
 \int_R \rd A = \int_{-\pi}^\pi a b\varepsilon \,\rd \varphi = 
 2\pi   a b \varepsilon
\end{equation}
and its volume is therefore $V_R=2\pi a b \varepsilon h $. This leads to the important result that 
\begin{equation}
 \frac{1}{A_R} \rd A =\frac{1}{\varrho V_R} \,\varrho h\,\rd A = \frac{1}{2\pi a b \varepsilon h} \, a b \varepsilon h\, \rd\varphi =\frac{1}{2\pi}\,\rd \varphi
\end{equation}
such that area averages, or mass-weighted area averages, over the elliptical ring can be expressed as phase averages over~$\varphi$. Such quantities are independent of the ellipse width $\varepsilon$, height $h$, and density $\varrho$.

At this point, we introduce a new notation. Let $f(\varphi,\bx,t)$ be some function of the free phase $\varphi$ and possibly also of space $\bx$ and time $t$. The \emph{phase average} of $f$ is defined by
\begin{equation}
\left\langle f(\varphi,\bx,t)\right\rangle
 \equiv \frac{1}{2\pi}
\int_{-\pi}^{\pi} f(\varphi,\bx,t)\,\rd \varphi
\end{equation}
an operation that gives the average value of $f$ experienced by all fluid parcels along the elliptical ring. The three integrals defined in Equation~(\ref{threeintegrals}) become, again using Equation~(\ref{crossproduct}) for cross products,
\begin{equation}\label{Iint}
I(t) = \left\langle \breve\bx^T\breve \bx\right\rangle,\quad\quad
M(t) = \left\langle \breve \bx^T\bJ^T\breve\bu\right\rangle,\quad\quad
\Pi(t) = \left\langle\left(\frac{\partial \breve\bx}{\partial \varphi}\right)^{\!T}\breve\bu\right\rangle
\end{equation}
where $\breve\bx$ and $\breve\bu$ in these expressions are the free-phase versions $\breve\bx(\varphi,t)$ and $\breve\bu(\varphi,t)$, respectively. For the moment of inertia, one finds upon substituting for $\breve\bx$ from Equation~(\ref{kinematicmodel})
\begin{equation}
I(t) \equiv \left\langle \breve\bx^T\!(\varphi,t)\,\breve \bx(\varphi,t)\right\rangle= 
 \left\langle \frac{a^2+b^2}{2} +\frac{a^2-b^2}{2}\cos2\varphi\right\rangle = \frac{a^2+b^2}{2}.
\end{equation}

In computing the angular momentum and circulation, it is convenient to define the two new vectors
\begin{equation}
\br \equiv \begin{bmatrix}
a \cos \varphi \\ b \sin \varphi\end{bmatrix},\quad\quad \bs \equiv \begin{bmatrix}
-a \sin \varphi \\ b \cos \varphi\end{bmatrix}
\end{equation}
such that $\breve\bx(\varphi,t)=\bR(\theta)\br$, while $\frac{\partial}{\partial \varphi}\breve\bx(\varphi,t)=\bR(\theta)\bs$. The integrands in the phase averages for $M(t)$ and $\Pi(t)$ become, respectively,
\begin{equation}\label{twointegrands}
\breve \bx^T\bJ^T\breve\bu = \br^T \left[\bR^T(\theta)\bJ^T\breve\bU \bR(\theta)\right]\br,\quad\quad \left(\frac{\partial \breve\bx}{\partial \varphi}\right)^{\!T}\breve\bu=\bs^T\left[ \bR^T(\theta) \breve \bU \bR(\theta)\right]\br
\end{equation}
where the matrices in square brackets are readily found, using Equation~(\ref{ellipseflowmatrix}), to be given by
\begin{align}\label{matrix1}
\bR^T(\theta)\bJ^T\breve\bU \bR(\theta)&=-\frac{\rd \ln \rho}{\rd t}\bJ +\left(\frac{\rd\theta}{\rd t}+\frac{\eta^2+1}{2\eta}\frac{\rd\phi}{\rd t}\right)\bI- \frac{1}{2}\frac{\rd\ln \eta}{\rd t}\bL-\frac{\eta^2-1}{2\eta}\frac{\rd\phi}{\rd t}\bK\\
\bR^T(\theta)\breve\bU \bR(\theta)&=\frac{\rd \ln \rho}{\rd t}\bI +\left(\frac{\rd\theta}{\rd t}+\frac{\eta^2+1}{2\eta}\frac{\rd\phi}{\rd t}\right)\bJ+ \frac{1}{2}\frac{\rd\ln \eta}{\rd t}\bK-\frac{\eta^2-1}{2\eta}\frac{\rd\phi}{\rd t}\bL.\label{matrix2}
\end{align}
The $\bI\bJ\bK\bL$-modified inner products between $\br$ and itself, and between $\br$ and $\bs$, involve trigonometric terms which will reduce after phase averaging. The necessary phase-averaged inner products are
\begin{align}\label{phaseavgrules}
\left\langle\br^T\br\right\rangle
 &=\frac{ a^2+b^2}{2},&
\left\langle\br^T\bJ\br\right\rangle &=0,&
\left\langle\br^T\bK\br\right\rangle & =\frac{a^2-b^2}{2},&
\left\langle\br^T\bL\br\right\rangle &=0\\
\left\langle \bs^T\br\right\rangle
 &=0,&
\left\langle\bs^T\bJ\br\right\rangle &=ab,&
\left\langle\bs^T\bK\br\right\rangle&=0,&
\left\langle\bs^T\bL\br\right\rangle &=0.
\end{align}
Combining these values with the appropriate coefficients from Equations~(\ref{matrix1}) and (\ref{matrix2}) gives the forms of the angular momentum and circulation presented earlier in Equation~(\ref{threeintegralvalues}).

At this point, we will also derive the relationship between disk integrals and ring integrals stated earlier in Section~\ref{diskintegrals}. We will integrate over concentric rings, with the location of each ring designated by a parameter $\ell$ that varies from zero to one, with $\ell=1$ corresponding to the outer boundary of the ellipse. The differential area is $\rd A = \|\left(\ell\breve \bx \,\rd \ell\right)\times \rd\breve \bx \|= a b \ell \,\rd \ell\rd\varphi $. If the height $h$ of the disk is assumed to be uniform, the volume of the disk is
\begin{equation}
V_D=\iint_{D} h \,\rd A = \int_{-\pi}^\pi \int_{0}^1 h a b \ell \,\rd \ell\rd\varphi = \pi a b h.
\end{equation}
The average of a quadratic quantity $\bx^T \bG \bx$ over the entire disk is then found to be
\begin{multline}
\frac{1}{A_D}\iint_D \,\bx^T \bG \bx \,\rd A = \frac{1}{\pi a b }\int_{-\pi}^\pi \int_{0}^1 \left(\ell\breve\bx\right)^T \bG \left(\ell\breve\bx\right) a b \ell \,\rd \ell\rd\varphi \\= \int_{0}^1 \ell^3\rd\ell \, \frac{1}{\pi}\int_{-\pi}^\pi \breve \bx^T\bG \breve\bx\,\rd\varphi =\frac{1}{4\pi}\int_{-\pi}^\pi \breve \bx^T\bG \breve\bx\,\rd\varphi =\frac{1}{2}\left\langle\breve \bx^T\bG \breve\bx\right\rangle
\end{multline}
showing that the average of $\bx^T \bG \bx$ over the entire disk is one half of its average in any ring, as claimed.

\subsection{Computing the Kinetic Energy Integral}

To derive an expression for the average kinetic energy of an elliptical ring, given earlier in Equation~(\ref{kexpression}), we write the kinetic energy integral as a phase average,
\begin{equation}
K(t)=\frac{1}{A_R}\iint_R \frac{1}{2} \|\bu\|^2 \rd A=\frac{1}{2}\left\langle \breve \bx^T \breve\bU^T\breve\bU \breve \bx\right\rangle,
\end{equation}
and then employ the expansion of the kinetic energy matrix from Equation~(\ref{kematrix}). Writing the ellipse flow matrix for convenience as $\breve\bU=\frac{1}{2}\bR^T(\theta)\left\{\delta\bI+\zeta\bJ+\tilde\gn\bK+\tilde \gs\bL\right\}\bR(\theta)
$, where $\tilde \gn$ and $\tilde\gs$ are the normal and shear strains in the reference frame of the ellipse, Equation~(\ref{kematrix}) becomes
\begin{equation}
\breve\bU^T\breve\bU= \frac{1}{4}\bR^T(\theta)\left\{\left(\delta^2+\zeta^2+\gamma^2\right)\bI + 2\left(\delta\tilde\gn+\zeta\tilde\gs\right)\bK + 2\left(\delta\tilde\gs-\zeta\tilde\gn\right)\bL\right\}\bR(\theta)
\end{equation}
for the kinetic energy matrix associated with the ellipse evolution. Then, from the phase averaging rules given in Equation~(\ref{phaseavgrules}), the kinetic energy is found to be 
\begin{equation}
K(t)=\frac{1}{8} \left\{\frac{a^2+b^2}{2}\left(\delta^2+\zeta^2+\gamma^2\right) 
+2\frac{a^2-b^2}{2}\left(\delta\tilde\gn+\zeta\tilde \gs\right)\right\}.
\end{equation}
It turns out that further manipulations will be considerably easier if we work with the ellipse extension $\mu$ rather than with the aspect ratio $\eta$, so we rewrite Equation~(\ref{ellipseflowmatrix}) for $\breve \bU$ as
\begin{equation}\label{ellipseflowmatrixmu}
\breve\bU(t)=\bR(\theta)\left\{
\frac{\rd \ln \rho}{\rd t}\bI +\left(\frac{\rd\theta}{\rd t}+\mu\frac{\rd\phi}{\rd t}\right)\bJ+\frac{1}{\sqrt{\mu^2-1}}\frac{1}{2}\frac{\rd\mu }{\rd t}\bK-\sqrt{\mu^2-1}\,\frac{\rd\phi}{\rd t}\bL\right\}\bR^T(\theta)
\end{equation}
again using Table~\ref{eccentricitytable} and the derivative relations in Equation~(\ref{derivrelations}). This leads at once to 
\begin{multline} 
K= \frac{1}{2}\rho^2 \mu\left[\left(\frac{\rd \ln \rho}{\rd t}\right)^2+
\left(\frac{\rd\theta}{\rd t}+\mu\frac{\rd\phi}{\rd t}\right)^2+
\frac{1}{\mu^2-1}\left(\frac{1}{2}\frac{\rd\mu }{\rd t}\right)^2+
\left(\mu^2-1\right)\left(\frac{\rd\phi}{\rd t}\right)^2
\right]\\
+\rho^2 \sqrt{\mu^2-1}\left[\frac{\rd \ln \rho}{\rd t}\frac{1}{\sqrt{\mu^2-1}}\frac{1}{2}\frac{\rd\mu }{\rd t}- \left(\frac{\rd\theta}{\rd t}+\mu\frac{\rd\phi}{\rd t}\right)\sqrt{\mu^2-1}\frac{\rd\phi}{\rd t}\right]
\end{multline}
which may be rearranged to give 
\begin{equation}
K= \frac{1}{2}\rho^2\mu\left\{ \left(\frac{\rd \ln \rho}{\rd t}+\frac{1}{2}\frac{\rd\ln \mu}{\rd t}\right)^2+
\frac{1}{\mu^2}\left(\frac{\rd\theta}{\rd t}+\mu\frac{\rd\phi}{\rd t}\right)^2
+\frac{\mu^2-1}{\mu^2}\left(\frac{\rd\theta}{\rd t}\right)^2+\frac{1}{\mu^2-1}\left(\frac{1}{2}\frac{\rd\ln\mu}{\rd t}\right)^2\right\}.
\end{equation}
Re-expressing this in $\eta$ form leads to Equation~(\ref{kexpression}), as claimed.


\subsection{A Partitioning of the Ellipse Kinetic Energy}

The average kinetic energy of an elliptical ring has a simple partitioning into three portions, as is now shown. With $\breve\varpi(\varphi,t)\equiv \bk \cdot \breve \bx(\varphi,t) \times \breve \bu(\varphi,t)$ being the instantaneous angular velocity at each point along the ellipse, parameterized by the free phase $\varphi$, we find that the kinetic energy takes the form
\begin{equation} \label{wewillshow}
K(t)= \frac{1}{2}\frac{1}{I}\left\{ \Pi^2+ \left(\frac{1}{2}\frac{\rd I}{\rd t}\right)^2
+2\left\langle\left(\breve\varpi-\left\langle\breve\varpi\right\rangle \right)^2\right\rangle\right\}
\end{equation}
where the first term is again due to the circulation, the second is due to the change in the moment of inertia, and the third is identified as the \emph{angular velocity variance} along the ellipse. This term is given by
\begin{equation}\label{angmomvar}
\left\langle\left(\breve\varpi-\left\langle\breve\varpi\right\rangle \right)^2\right\rangle
=\frac{1}{2}I^2\left\{\lambda^2\left(\frac{\rd\theta}{\rd t}\right)^2+\frac{1}{1-\lambda^2}\left(\frac{1}{2}\frac{\rd\lambda}{\rd t}\right)\right\}
\end{equation}
and thus involves contributions from both precession and deformation. When fluid parcels at various phase locations around the elliptical annulus all have the same angular velocity, this term vanishes. 

 

To prove Equation~(\ref{angmomvar}), we note that the angular velocity along the ellipse periphery can be expressed as $\breve\varpi(\varphi,t)=\breve\bx^T\bJ^T\breve\bU\breve\bx$, and thus
\begin{equation}
\breve\varpi(\varphi,t)-\left\langle\breve\varpi\right\rangle 
=\breve\bx^T\bJ^T\breve\bU\breve\bx - \left\langle \breve\bx^T\bJ^T\breve\bU\breve\bx\right\rangle
\end{equation}
gives the deviation of this quantity from its phase mean. The quadratic form $\breve\bx^T\bJ^T\breve\bU\breve\bx$ was previously examined in Equations~(\ref{twointegrands}) and (\ref{matrix1}), but unlike in that case, we will now also need to keep track of terms dependent on $\varphi$. The following combinations involving $\br$ will occur in $\breve\bx^T\bJ^T\breve\bU\breve\bx$:
\begin{align}\label{phaseaveragedterms1}
\br^T\bI\br & = \frac{a^2+b^2}{2} + \frac{a^2-b^2}{2}\cos2\varphi, & 
\br^T\bJ\br & = 0, \\
\br^T\bK\br& =\frac{a^2-b^2}{2} + \frac{a^2+b^2}{2}\cos2\varphi,&
\br^T\bL\br & =ab\sin2\varphi. \label{phaseaveragedterms2}
\end{align}
However, only the sinusoidal terms need be considered, as the constant terms are removed by subtracting the phase average. One finds that the deviation of the angular velocity from its phase-averaged value is
\begin{equation}
\breve\varpi(\varphi,t)-\left\langle\breve\varpi\right\rangle 
=ab\left\{\frac{\eta^2-1}{2\eta} \frac{\rd\theta}{\rd t} \cos 2\varphi - \frac{1}{2} \frac{\rd\ln\eta}{\rd t} \sin 2\varphi \right\}
\end{equation}
after making use of Equation~(\ref{matrix1}), and noting a cancellation of all terms involving the orbital frequency $\frac{\rd}{\rd t}\phi(t)$. Squaring this expression and applying the phase average again then leads to 
\begin{equation}
\left\langle\left(\breve\varpi-\left\langle\breve\varpi\right\rangle \right)^2\right\rangle
= \frac{1}{2} a^2b^2 \left\{\left(\frac{\eta^2-1}{2\eta}\right)^2\left( \frac{\rd\theta}{\rd t}\right)^2+ \left(\frac{1}{2} \frac{\rd\ln\eta}{\rd t}\right)^2 \right\}
\end{equation}
and transforming from the aspect ratio $\eta$ to the linearity $\lambda$, we obtain the form given in Equation~(\ref{angmomvar}).

\section{The Extended Stokes' Theorem}\label{stokes}

In this section, it is shown that the principle of ellipse/flow equivalence can be understood as an extended version of Stokes' theorem that expresses relationships between contour and area integrals involving all four terms in the velocity gradient matrix. 

\subsection{Moment Matrices}

The physical properties encountered in the previous section can be compactly combined into two matrix-valued quantities, defined as
\begin{equation}
\bM(t)\equiv \frac{1}{A_R} \iint_R \,\bx\bx^T \rd A,\quad\quad
\bN(t) \equiv\frac{1}{2\pi} \oint_C \bu\,\rd\bx^T \label{MandNdef}
\end{equation}
which, with $R$ again denoting the elliptical ring and $C$ its boundary, can be expressed as
\begin{equation}
\bM(t) = \frac{1}{2\pi}\int_{-\pi}^\pi \breve\bx\breve\bx^T \rd \varphi=
\left\langle\breve \bx \breve \bx^T \right\rangle,\quad\quad
\bN(t) = \frac{1}{2\pi}\int_{-\pi}^\pi\breve\bu\,\frac{\partial\breve\bx^T}{\partial \varphi} \rd \varphi =
\left\langle\breve \bu \,\frac{\partial }{\partial\varphi}\breve\bx^T \right\rangle
\end{equation}
in terms of phase averages. The \emph{moment matrix} $\bM(t)$ for the elliptical annulus is found to be given by 
\begin{equation}
\bM(t)= 
 \frac{1}{2} \bR(\theta) \bD\bR^T(\theta),\quad\quad\bD = \begin{bmatrix}
 a^2 & 0 \\ 0 & b^2
\end{bmatrix}
\end{equation}
and as such is recognized as one half of the inverse of the ellipse matrix $\bE(t)$ encountered earlier, $\bM=\frac{1}{2}\bE^{-1}$. Under the action of advection by a linear flow $\bu=\bU\bx$, the ellipse matrix $\bE$ and moment matrix $\bM$ evolve according to the formally different, but equivalent, laws:
\begin{equation}\label{EandMevolutions}
\frac{\rd \bE}{\rd t} = -\bE\bU - \bU^T \bE,\quad\quad\frac{\rd \bM}{\rd t} = \bU\bM + \bM \bU^T.
\end{equation}
The first of these was derived earlier in Section~\ref{kinematicsection}, while second is readily found by differentiating $\bE\bM=\frac{1}{2}\bI$, then substituting the first equation; see p. 152 of \cite{mckiver03-jfm} or p. 5 of \cite{mckiver15-amp}, who~derive these two equations in the context of an ellipsoid in a three-dimensional linear flow. 

The second matrix, $\bN(t)$, will be referred to as the \emph{circulation matrix}, as we recognize from its definition in Equation~(\ref{MandNdef}) that $\tr\{\bN\}=\Pi$. Because $\bN(t)$ is defined as a contour integral, not a spatial average, it is a property of the elliptical \emph{boundary} rather than of the \emph{annulus}. The circulation matrix for the ellipse is found to be
\begin{equation}
\bN(t)= \left\langle\breve \bu \frac{\partial}{\partial\varphi}\breve\bx^T \right\rangle=\breve\bU \bR(\theta)\left\langle \br\bs^T \right\rangle \bR^T(\theta) = \frac{1}{2} ab \breve\bU \bJ^T
\end{equation}
where $\br=[a\cos\varphi\,\, \,\,b\sin \varphi]^T$ and $\bs=[-a\sin\varphi\,\,\,\, b\cos \varphi]^T$ as defined previously, and where the final equality follows from observing that $\left\langle \br\bs^T \right\rangle = \frac{1}{2} ab \bJ^T$. Substituting from the definition of $\breve \bU(t)$ given in Equation~(\ref{ellipseflowmatrix}) leads to 
\begin{equation}
\bN(t)= \frac{1}{2} ab \bR(\theta)\left\{
\left(\frac{\rd\theta}{\rd t}+\frac{\eta^2+1}{2\eta}\frac{\rd\phi}{\rd t}\right)\bI-\frac{\rd \ln \rho}{\rd t}\bJ +\frac{\eta^2-1}{2\eta}\frac{\rd\phi}{\rd t}\bK+ \frac{1}{2}\frac{\rd\ln \eta}{\rd t}\bL\right\}\bR^T(\theta).
\end{equation}
In the next section, it is established that there is an extension of the classical Stokes' theorem that lets us evaluate the matrix $\bN(t)$ as an area integral, rather than as a contour integral. Using that result, we find that $\bN$
can be alternately expressed as 
\begin{equation}
\bN(t) = \label{Nexpression}
\frac{1}{4} ab \bR(\theta)\left[\zeta\bI - \delta\bJ -\gamma \sin 2\tilde \alpha \bK +\gamma \cos\ 2\tilde \alpha \bL \right]\bR^T(\theta)
\end{equation}
and equating coefficients of the $\bI\bJ\bK\bL$ matrices between the previous two equations, we obtain Equation~(\ref{correspond}) for the correspondence between the velocity gradient quantities and the ellipse rates of change. Thus the principle of ellipse/flow equivalence is actually a manifestation of Stokes' theorem, for the particular case of a linear flow and with the region of integration chosen as an ellipse.

\subsection{An Extended Stokes' Theorem}

The generalized form of Stokes' theorem relates the contour integral of a quantity to a spatial integral of a derivative of that quantity. Its two most common manifestations are the Kelvin--Stokes or classical Stokes' theorem, relating vorticity to circulation, and the divergence theorem, relating outward flux to divergence. Similar integral relations may equally well be created for the two components of the strain field, although the author is not familiar with any literature in which these appear.


In this section, a version of Stokes' theorem will be proven that accommodates all four components of the velocity gradient matrix. Let $A$ be some area in a flat two-dimensional domain bounded by the curve $C$, with $\rd \bx$ being a differential element of $C$. We form the $2\times 2$ matrix $\bu\,\rd\bx^T$ as the \emph{outer product} of the velocity and the differential element of the boundary. This matrix is expanded as
\begin{equation}\label{udx}
\bu\,\rd\bx^T = \frac{1}{2} \left\{
\left(\bu^T\rd\bx\right)\bI
+\left(\bu^T\bJ\,\rd\bx\right)\bJ
+\left(\bu^T\bK\,\rd\bx\right)\bK
+\left(\bu^T\bL\,\rd\bx\right)\bL
\right\} 
\end{equation}
which follows from Equation~(\ref{jklexpansion}), substituting $\bu\,\rd\bx^T$ for $\bU$ and then noting that the coefficients of the $\bI\bJ\bK\bL$ matrices take the forms $\tr\left\{\bu\,\rd\bx^T\bG^T\right\}=\bu^T\bG\,\rd\bx$. We can choose to think of these coefficients either as inner products between $\rd\bx$ and modified versions of the flow field $\bu$, or as inner products between $\bu$ and modified versions of the differential element $\rd\bx$. 

There is a simple relationship between the contour integral of the matrix $\bu\,\rd\bx^T$ and the spatial integral of the velocity gradient matrix:
\begin{equation}\label{geometric}
\oint_C \bu\,\rd\bx^T = \iint_A 
\left(\bm{\nabla}\bu^T\right)^T \bJ^T\rd A.
\end{equation}
This relationship, which will be called the \emph{extended Stokes' theorem}, expands to give, using~Equation~(\ref{uijkl}) for the right-hand side,
\begin{equation}
\label{stokesexpanded}
\begin{aligned}
\oint_C \bu\,\rd\bx^T&=\frac{1}{2}\left\{\left(\oint_C \bu^T\rd\bx\right) \bI + \left(\oint_C \bu^T\bJ\rd\bx\right) \bJ+\left(\oint_C \bu^T\bK\,\rd\bx\right) \bK + \left(\oint_C \bu^T\bL\,\rd\bx\right) \bL\right\}\\
&=\frac{1}{2}\iint_A \left\{
\left(\bm{\nabla}^T \bJ^T\bu \right)\bI
-\left( \bm{\nabla}^T \bu\right)\bJ
-\left(\bm{\nabla}^T \bL\bu\right)\bK
+\left(\bm{\nabla}^T \bK\bu\right)\bL
\right\} \rd A \\
&=\frac{1}{2} \,\iint_A \left\{
\zeta\bI
-\delta\bJ
-\gs\bK
+\gn\bL
\right\} \rd A \equiv \frac{1}{2} A \left\{\overline\zeta\bI
-\overline\delta\bJ
-\overline\gs\bK
+\overline\gn\bL\right\}
\end{aligned}
\end{equation}
which is seen to encapsulate four separate integral relations into its four matrix components. The four overlined quantities on the right-hand side---$\overline \zeta$, $\overline \delta$, $\overline \sigma$, and $\overline \nu$---are defined to be spatial averages, with $A$ in this expression indicating the area of the corresponding region. Separating the four distinct relations in Equation~(\ref{stokesexpanded}) for greater clarity, we find that the $\bI$, $\bJ$, $\bK$, and $\bL$ components give, respectively,
\begin{align}
(\bI)&&\hspace{-.2in}\oint_C \bu^T \rd \bx & = \iint_A \bm{\nabla}^T\bJ^T \bu \,\rd A =A\overline \zeta\,& \oint_C \bu \cdot \rd\bx &= \iint_A \bk \cdot \left( \nabla \times \bu \right)\,\rd A=A\overline \zeta\\
(\bJ)&&\hspace{-.2in}\oint_C \bu^T \bJ^T\rd \bx &= \iint_A \bm{\nabla}^T \bu \,\rd A=A\overline \delta, & \oint_C \bu\cdot \left(\rd\bx \times \bk\right) &= \iint_A \nabla \cdot \bu \,\rd A= A\overline \delta\\
(\bK)&&\hspace{-.2in}-\oint_C \bu^T \bK\,\rd \bx &= \iint_A \bm{\nabla}^T \bL \bu \,\rd A= A\overline \sigma,
& -\oint_C \bK\bu \cdot\rd \bx & =\iint_A \nabla \cdot \bL\bu \,\rd A = A\overline \sigma \\
(\bL)&&\hspace{-.2in}\oint_C \bu^T \bL\,\rd \bx &= \iint_A \bm{\nabla}^T \bK\bu \,\rd A= A\overline \nu,
&\oint_C \bL\bu \cdot\rd \bx & =\iint_A \nabla \cdot \bK\bu \,\rd A = A\overline \nu
\end{align}
which are written with our matrix-based notation on the left and standard notation on the right. The~first two relationships are recognized as the classical Stokes' theorem and the divergence theorem, respectively; note that the $\bJ$ and $\bK$ relations appear in Equation~(\ref{stokesexpanded}) multiplied by negative one from the way they are written here. For the divergence theorem, note that $\bJ\bu\cdot \rd \bx=\bu^T\bJ^T\rd\bx=\bu \cdot \bJ^T\bx=\bu\cdot \left(\rd\bx \times \bk\right)$, where the vector $\rd\bx \times \bk\equiv \rd\bn$ is recognized as the exterior normal, and also that $\nabla\times(\bJ\bu)=\bm{\nabla}^T\bJ^T\bJ\bu=\bm{\nabla}^T\bu=\nabla\cdot\bu$.  The third and fourth relations, which we may call the \emph{reflectional Stokes' theorems}, link the \emph{spatially-integrated strain} to integrals of velocity along the region boundary. 

The four components of the extended Stokes' theorem link the $\bI\bJ\bK\bL$-modified inner products, appearing in Equation~(\ref{udx}), with the $\bI\bJ\bK\bL$-modified gradient operators that were seen in Equation~(\ref{uijkl}) to occur in $\left(\bm{\nabla}\bu^T\right)^T$. Observe, however, that the matrices appearing in the contours integrals and~those appearing in the area integrals are not the same; this is due to the $\bJ^T$ matrix in Equation~(\ref{geometric}). This~leads in particular to an unfamiliar appearance for the strain relations, the first of which involves $-\bK$ on the left and $\bL$ on the right, and the second of which involves $\bL$ on the left and $\bK$ on the right. 

An alternate presentation of the extended Stokes' theorem is, with $\rd\bn\equiv \bJ^T\rd\bx=\rd\bx \times \bk$ being the exterior normal,
\begin{equation}\label{geometricvariant}
 \oint_C \bu\,\rd\bn^T = \iint_A 
\left(\bm{\nabla}\bu^T\right)^T \rd A.
\end{equation}
This has the advantage that the matrices appearing on the contour integral side and the area integral side are the same. However, one may view 
$\bu\, \rd\bx^T $ as a more natural quantity than $\bu\,\rd\bn^T $, as~the differential element $\rd \bx$ appears to describe the curve $C$ more directly than does the differential normal $\rd \bn$. Moreover, the matrix $\bu\, \rd\bx^T$ places the vorticity, the most important physical quantity for most fluid dynamics problems, in the most prominent location, i.e., along the trace, while $\bu\, \rd\bn^T$ has divergence along the trace. Normally, one writes the circulation integral in terms of the differential element $\rd \bx$ and the contour integral for the divergence theorem in terms of the differential normal $\rd \bn$; here, however, one must make a choice. As the two versions of the extended Stokes' theorem contain equivalent information, which version one uses can be determined by personal preference. 



The extended Stokes' theorem may be proven as follows. For some area $A$ bounded by a~contour~$C$, the classical Stokes' theorem states that the circulation is equal to the enclosed vorticity:
\begin{equation}
\oint_C \bu^T \rd \bx = \iint_A \bm{\nabla}^T\bJ^T \bu \,\rd A,\quad\quad \oint_C \bu \cdot \rd\bx = \iint_A \bk \cdot \left( \nabla \times \bu \right)\,\rd A.
\end{equation}
This relationship appears as the trace of Equation~(\ref{stokesexpanded}), due to the fact that $\tr\left\{\bu\,\rd\bx^T\right\}=\bu^T\rd\bx$ while $\tr\left\{\left(\bm{\nabla}\bu^T\right)^T\bJ^T\right\}=\bm\nabla^T\bJ^T\bu=\frac{\partial}{\partial x}v-\frac{\partial}{\partial y} u$. However, in this theorem, one can replace $\bu$ with a modified version, say $\bG^T\bu$ for some matrix $\bG$, to obtain the modified version
\begin{equation}
\oint_C \bu^T\bG \rd \bx = \iint_A \bm{\nabla}^T\bJ^T \bG^T\bu \,\rd A,\quad\quad \oint_C \bG^T\bu \cdot \rd\bx = \iint_A \bk \cdot \left( \nabla \times \bG^T\bu \right)\,\rd A.\label{stokesproof}
\end{equation} 
Now, multiplying both sides of Equation~(\ref{geometric}) from the right by $\bG^T$ and taking the trace, one~obtains Equation~(\ref{stokesproof}); this is so because $\tr\left\{\bu\,\rd\bx^T\bG^T\right\}=\bu^T\bG\,\rd\bx$ on the left-hand side of Equation~(\ref{geometric}), while on the right-hand side, we have $\tr\left\{\left(\bm{\nabla}\bu^T\right)^T\bJ^T\bG^T\right\}=\bm\nabla^T\bJ^T\bG^T\bu$, as may be readily verified. Thus, choosing $\bG^T$ successively as each of $\bI$, $\bJ$, $\bK$, and $\bL$ in Equation~(\ref{stokesproof}) proves the corresponding component equation within Equation~(\ref{stokesexpanded}), which are then gathered together into Equation~(\ref{geometric}).




Returning to the case of an ellipse in a linear flow, we find that the circulation matrix $\bN$ becomes, on account of the extended Stokes' theorem,
\begin{equation}
2\pi \bN = \oint_C \bu\,\rd\bx^T =\int_{-\pi}^\pi \bu\,\frac{\partial }{\partial \varphi}\bx^T \rd \varphi= \iint_A 
\left(\bm{\nabla}\bu^T\right)^T \bJ^T 
\rd A =\frac{1}{2}\pi ab \left\{\zeta\bI
-\delta\bJ
-\gs\bK
+\gn\bL\right\}
\end{equation}
recovering the earlier expression for $\bN$ given in Equation~(\ref{Nexpression}) and establishing that ellipse/flow equivalence is a manifestation of Stokes' theorem.

It should be emphasized that the extended Stokes' theorem is not fundamentally new; it is simply the classical Stokes' theorem, or equivalently the divergence theorem, applied to four different vector fields, the original velocity field together with three modified versions thereof. However, in the same sense, the divergence theorem is not fundamentally different from Stokes' theorem either. What is new here is the extension or generalization of those results to establish a fundamental relationship between a velocity field and all four components of its gradient, united into one simple equation.




\section{Discussion}

This paper has examined basic properties of a passive fluid ellipse advected by a linear flow. While this problem would seem to be elementary, it turns out to be surprisingly rich. By approaching the problem with a parametric model for the motion of a fluid particle in an ellipse, together with the use of a new matrix-based notation, several new results have been obtained. 

The main result is the equivalence between the Lagrangian perspective of an evolving ellipse and the Eulerian perspective of a velocity field that depends linearly on the spatial coordinates. This result, termed \emph{ellipse/flow equivalence}, has important implications for our understanding of how spatial information may be extracted from Lagrangian measurements, for example, in the study of coherent eddies. In particular, it is natural to ask under what conditions ellipses diagnosed by Lagrangian analysis methods such as that of Lilly et al. \cite{lilly11-grl} correspond to physical structures in the flow and, therefore, encode in principle information about all local gradients of the velocity field. Moreover, it establishes that an evolving ellipse is fundamental to a Lagrangian perspective in the same way that a linear approximation to a velocity field is fundamental to an Eulerian perspective.

Ellipse/flow equivalence was shown to be a manifestation of a more general result, called the extended Stokes' theorem. This theorem shows how the contour integral of a simple matrix-valued quantity along the boundary of any region recovers the spatial average of all four elements of the velocity gradient tensor within that region. This result incorporates integral relations regarding the two components of the strain field, relations that are rarely encountered. A main strength of this result is its notational compactness, combining four different relationships into one simple equation.

Expressions for integrated physical properties of a fluid ellipse were derived using the parametric model. A new expression was given for the average kinetic energy along an elliptical ring, which was shown to have a partitioning into three distinct portions, associated respectively with the circulation, the rate of change of the moment of inertia, and finally a simple quantity, which is nevertheless rarely encountered, the \emph{variance} of the angular velocity. This kinematic result was shown to have an intriguing similarity to a dynamical equation for the evolution of the moment of inertia in shallow-water elliptical vortices \cite{ball63-jfm,young86-jfm,holm91-jfm}, a comparison that calls for a more thorough investigation. 


There are several obvious directions for future research. The first is to use these results to investigate solutions for Kida-type vortices, as well as for shallow-water elliptical vortices, work that is currently underway. These results are also relevant to ellipsoidal vortices in three dimensions \cite{meacham92-dao,mckiver03-jfm,mckiver06-jfm,mckiver15-amp}. In those studies, a vortex is embedded in a two-dimensional linear flow that is allowed to vary with height. As such, the ellipsoid evolves in a layer-wise fashion, with the ellipse within each layer advected by a different linear flow; therefore, this work applies here just as it does to two-dimensional vortices. Furthermore, the results herein point to a deeper and more direct connection between Lagrangian and Eulerian properties than may commonly be appreciated. Understanding that connection more fully may help to extract more information from Lagrangian data, and may also be relevant to active research in the field of Lagrangian coherent structures (e.g., \cite{beron-vera08-grl,rutherford12-acp,haller12-pd}).

Herein, we have assumed that the parametric model accurately describes the evolution of an~initial elliptical ring of fluid parcels. If the flow is linear, this will be the case. However, if the flow is not linear, there is no reason to suppose that a material curve that begins as an ellipse will remain as an ellipse. One would expect non-linear aspects of the velocity field (that is, higher-order terms in a~local Taylor series expansion) to introduce perturbations to the elliptical shape. The parametric model would therefore require modifications in situations in which the flow is not linear. 

An interesting direction would be to quantify the extent to which non-linear aspects of the velocity field correspond to deviations from elliptical evolution. That is, if a flow is \emph{approximated} as being locally linear, to what extent can the nearby Lagrangian behavior be approximated as being such that ellipses remain ellipses? The extended Stokes' theorem points to a correspondence between these two approximations, in the following sense. Take the contour integral over an elliptical region within which the flow is not necessarily linear. From the extended Stokes' theorem, one obtains \emph{average} values of the divergence, vorticity, and strain terms. Naturally, one conjectures that the initially elliptical boundary will evolve primarily as if it were within a \emph{linear} flow characterized by the average values of the enclosed \emph{non-linear} flow, together with a perturbation to this elliptical evolution from higher-order components of the flow field. Formalizing this approximation would provide a still better understanding of the local connection between the Eulerian and Lagrangian perspectives. 

At the same time, it would be valuable to investigate the ellipse/flow correspondence for cases in which the ellipse is not an actual material ellipse, but rather the moments of a distribution. It~is commonplace to attempt to describe an oceanic tracer distribution, or a vorticity distribution in a~numerical model, in terms of its area moments. This essentially describes a distribution \emph{as if it were} an~ellipse, leading to a matrix similar to what is called the `moment matrix' here. To what extent will that distribution also evolve as if it were an ellipse, with its evolution dictated by the average properties of the enclosed velocity field? Again, one would expect, based on the extended Stokes' theorem, that there may be a close connection between the evolution of an actual ellipse and that of a `moment~ellipse', and that there may be an avenue for examining this connection in a~rigorous~manner. 

For applications of stirring and mixing, it may be that the higher-order perturbations are actually of greater interest, because these may dominate deformation of the boundary and therefore the potential for small-scale exchange and diffusion. Because evolution under a locally linear, time-dependent flow maps an ellipse into another ellipse, it is reversible and therefore cannot by itself account for mixing. For such problems, this work may be helpful in isolating the smoothly-deforming, reversible portion of the flow from those processes driving higher-order deformation. 

Finally, as pointed out by an anonymous reviewer, the classical Stokes' theorem applies to any curved surface in three dimensions, whereas the assumption of a flat domain has been used in the derivation of the extended Stokes' theorem. It seems intuitive that the extended Stokes' theorem would have an extension to curved surfaces, for which one would need to correctly account for metric terms arising from the curvature. Such a result could have applications in planetary fluid dynamics, which are often idealized as two-dimensional or shallow-water flows on the surface of a sphere. 

\vspace{6pt} 


\paragraph{Acknowledgements}  This work was support by United States National Science Foundation awards \#1235310 and \#1459347.  The author is grateful to Darryl Holm and Jeffrey Early for helpful and inspiring conversations, as well as to two anonymous reviewers for their helpful comments.\\

\appendix
  
\noindent{\bf Appendix: The Kinematic Boundary Condition}\vspace{1mm}

Here we verify that the kinematic condition for the motion of an elliptical boundary, Equation~(\ref{kinematic}), recovers the first three ellipse evolution equations presented in Equation~(\ref{etol}). Whereas \citet{neu84-pf} and \citet{ide95-fdr} write out the entries of the matrix emerging from the derivative of the boundary condition in terms of the rates of change of the ellipse parameters, we will show how the ellipse rates of change express themselves in the $\bI\bJ\bK\bL$ basis decomposition of that matrix. 


The matrix equation derived in the main text from the kinematic boundary condition for the evolution of $\bE$, Equation~(\ref{matrixequation}), can be rearranged to become
\begin{equation}\label{modifiedmatrix}
-\bR^T(\theta)\left\{\bE^{-1}\frac{\rd \bE}{\rd t}\right\}\bR(\theta) = 
\bR^T(\theta)\left\{\bE^{-1} \bU^T \bE + \bU\right\}\bR(\theta)
\end{equation}
which, for later convenience, has been rotated into the reference frame of the ellipse. With the ellipse matrix $\bE(t)$ defined as $\bE(t)\equiv \bR(\theta)\bD^{-1}\bR^T(\theta)$ as in Equation~(\ref{ellipsematrix}), its derivative is found to be
\begin{equation}\label{ederivative}
\frac{\rd \bE}{\rd t} = \left(\bJ\bE+\bE\bJ^T\right)\frac{\rd\theta}{\rd t}+
\bR(\theta)\frac{\rd\bD^{-1}}{\rd t}\bR^T(\theta),\quad\quad
\frac{\rd \bD^{-1}}{\rd t} = 
-\bD^{-1}\left[ \bI \frac{\rd \ln \rho^2}{\rd t} + \bK\frac{\rd \ln \eta}{\rd t}\right]
\end{equation}
making use of $a=\rho\sqrt{\eta}$ and $b=\rho/\sqrt{\eta}$ for the latter equation. Using the last two equations, the~left-hand side of Equation~(\ref{modifiedmatrix}) then becomes 
\begin{equation}\label{lhs}
-\bR^T(\theta)\left\{\bE^{-1}\frac{\rd \bE}{\rd t} \right\}\bR(\theta) = 
 \bI \frac{\rd \ln \rho^2}{\rd t} -\left[\bD\bJ\bD^{-1}-\bJ\right] \frac{\rd \theta}{\rd t}+ \bK\frac{\rd \ln \eta}{\rd t}.
\end{equation} 
Next, we rewrite the flow matrix $\bU(t)$ by expressing it in a reference frame aligned with the instantaneous ellipse orientation $\theta$, leading to
\begin{equation}\label{Udecompxxxx}
\bU(t) =
\frac{1}{2}\bR(\theta)\left[\delta \bI + \zeta \bJ+\gamma \cos2\tilde\alpha\bK +\gamma \sin2\tilde\alpha\bL\right]
\bR^T(\theta)
\end{equation}
where $\widetilde \alpha\equiv \alpha-\theta$ as in the main text. The right-hand side of Equation~(\ref{modifiedmatrix}) is then
\begin{multline}
\bR^T(\theta)\left\{\bE^{-1}\bU^T\bE +\bU \right\}\bR(\theta)= \frac{1}{2}\bD\left[ \delta \bI - \zeta \bJ+\gamma \cos2\tilde\alpha\bK +\gamma \sin2\tilde\alpha\bL\right]\bD^{-1}+\bR^T(\theta)\bU\bR(\theta)\\
=\delta\bI -\frac{1}{2}\left[\bD\bJ\bD^{-1}-\bJ\right] \zeta +\frac{1}{2}\left[\bD\bK\bD^{-1} +\bK\right]\gamma \cos2\tilde\alpha +\frac{1}{2}\left[\bD\bL\bD^{-1} +\bL\right]\gamma\sin2\tilde\alpha 
\end{multline}
and to simplify this, we may note the following, readily verifiable identities:
\begin{align}\label{jidentity}
 \bD\bJ\bD^{-1}-\bJ &
=\frac{\eta^2-1}{2\eta^2} \left[\left(\bJ-\bL\right)\eta^2 -\left(\bJ+\bL\right)\right] \\
 \bD\bK\bD^{-1}+\bK&= 2\bK\\ 
\bD\bL\bD^{-1}+\bL &
=-\frac{\eta^2+1}{2\eta^2} \left[\left(\bJ-\bL\right)\eta^2 -\left(\bJ+\bL\right)\right] \label{lidentity}.
\end{align}
The first and third of these combine to eliminate $\left[\bD\bL\bD^{-1}+\bL\right]$ in favor of $\left[\bD\bJ\bD^{-1}-\bJ\right]$, leading~to
\begin{equation}
\bR^T(\theta)\left\{\bE^{-1}\bU^T\bE +\bU \right\}\bR(\theta)
=\delta\bI -\frac{1}{2}\left[\bD\bJ\bD^{-1}-\bJ\right]\left(\zeta +\frac{\eta^2+1}{\eta^2-1}\gamma\sin2\tilde\alpha \right) +\bK\gamma \cos2\tilde\alpha \label{rhs}
\end{equation}
and equating the coefficients of the $\bI$, $\bK$ and $\left[\bD\bJ\bD^{-1}-\bJ\right]$ matrices between \mbox{Equations~(\ref{lhs}) and (\ref{rhs})}, we recover the first three evolution equations given in Equation~(\ref{etol}). Note that as $\left[\bD\bJ\bD^{-1}-\bJ\right]$ contributes only to the $\bJ$ and $\bL$ components, it does not interfere with the $\bI$ or $\bK$ components and~therefore need not be written out explicitly. This also means that the $\bJ$ and $\bL$ components contain equivalent information, resulting in only three equations, rather than four. 


\begin{thebibliography}{999}
\providecommand{\natexlab}[1]{#1}

\bibitem[Kida(1981)]{kida81-jpsj}
Kida, S.
\newblock Motion of an elliptic vortex in a uniform shear flow.
\newblock {\em J. Phys. Soc. Jpn.} {\bf 1981}, {\em 50},~3517--3520.

\bibitem[Neu(1984)]{neu84-pf}
Neu, J.C.
\newblock The dynamics of a columnar vortex in an imposed strain.
\newblock {\em Phys. Fluids} {\bf 1984}, {\em 27},~2397--2402.

\bibitem[Ide and Wiggins(1995)]{ide95-fdr}
Ide, K.; Wiggins, S.
\newblock The dynamics of elliptically shaped regions of uniform vorticity in
 time-periodic, linear external velocity fields.
\newblock {\em Fluid Dyn. Res.} {\bf 1995}, {\em 15},~205--235.

\bibitem[Dritschel(1990)]{dritschel90-jfm}
Dritschel, D.G.
\newblock The stability of elliptical vortices in an external straining flow.
\newblock {\em J. Fluid Mech.} {\bf 1990}, {\em 210}, 223--261.

\bibitem[Meacham and Flierl(1990)]{meacham90-dao}
Meacham, S.P.; Flierl, G.R.
\newblock Vortices in shear.
\newblock {\em Dyn. Atmos. Oceans} {\bf 1990}, {\em 14},~333--386.

\bibitem[Bayly \em{et~al.}(1996)Bayly, Holm, and Lifschitz]{bayly96-ptrsla}
Bayly, B.J.; Holm, D.D.; Lifschitz, A.
\newblock Three-dimensional stability of elliptical vortex columns in external
 strain flow.
\newblock {\em Philos. Trans. R. Soc. A} {\bf 1996}, {\em 354},~895--926.

\bibitem[Mitchell and Rossi(2008)]{mitchell08-pf}
Mitchell, T.B.; Rossi, L.F.
\newblock The evolution of {K}irchhoff elliptic vortices.
\newblock {\em Phys. Fluids} {\bf 2008}, {20}, 054103.

\bibitem[Guha \em{et~al.}(2013)Guha, Rahmani, and Lawrence]{guha13-pre}
Guha, A.; Rahmani, M.; Lawrence, G.A.
\newblock Evolution of a barotropic shear layer into elliptical vortices.
\newblock {\mbox{\em Phys. Rev. E}} {\bf 2013}, {\em 87}, 013020.

\bibitem[Koshel and Ryzhov(2017)]{koshel17-npg}
Koshel, K.V.; Ryzhov, E.A.
\newblock Parametric resonance in the dynamics of an elliptic vortex in a
 periodically strained environment.
\newblock {\em Nonlinear Process. Geophys.} {\bf 2017}, {\em 24},~1--8.

\bibitem[Bertozzi(1988)]{bertozzi88-sjma}
Bertozzi, A.L.
\newblock Heteroclinic orbits and chaotic dynamics in planar fluid flows.
\newblock {\em SIAM J. Math. Anal.} {\bf 1988}, {\em 19},~1271--1294.

\bibitem[Polivani \em{et~al.}(1990)Polivani, Wisdom, DeJong, and
 Ingersoll]{polvani90-science}
Polivani, L.M.; Wisdom, J.; DeJong, E.; Ingersoll, A.P.
\newblock Simple dynamical models of {N}eptune's great dark spot.
\newblock {\em Science} {\bf 1990}, {249}, 1393--1398.

\bibitem[Koshel \em{et~al.}(2013)Koshel, Ryzhov, and Zhmur]{koshel13-npg}
Koshel, K.V.; Ryzhov, E.A.; Zhmur, V.V.
\newblock Diffusion-affected passive scalar transport in an ellipsoidal vortex
 in a shear flow.
\newblock {\em Nonlinear Process. Geophys.} {\bf 2013}, {\em 20},~437--444.

\bibitem[Ngan \em{et~al.}(1996)Ngan, Meacham, and Morrison]{ngan96-pf}
Ngan, K.; Meacham, S.; Morrison, P.J.
\newblock Elliptical vortices in shear: {H}amiltonian moment formulation and
 {M}elnikov analysis.
\newblock {\em Phys. Fluids} {\bf 1996}, {\em 8},~896--913.

\bibitem[Vanneste and Young(2010)]{vanneste10-pf}
Vanneste, J.; Young, W.R.
\newblock On the energy of elliptical vortices.
\newblock {\em Phys. Fluids} {\bf 2010}, {\em 22},~081701.

\bibitem[Crosby \em{et~al.}(2013)Crosby, Johnson, and Morrison]{crosby13-pf}
Crosby, A.; Johnson, E.R.; Morrison, P.J.
\newblock Deformation of vortex patches by boundaries.
\newblock {\em Phys. Fluids} {\bf 2013}, {\em 25},~023602.

\bibitem[Melander \em{et~al.}(1986)Melander, Zabusky, and
 Styczek]{melander86-jfm}
Melander, M.V.; Zabusky, N.J.; Styczek, A.S.
\newblock A moment model for vortex interactions of the two-dimensional {E}uler
 equations. {P}art 1. {C}omputational validation of a {H}amiltonian elliptical
 representation.
\newblock {\em J. Fluid Mech.} {\bf 1986}, {\em 167},~95--115.

\bibitem[Legras and Dritschel(1991)]{legras91-pfa}
Legras, B.; Dritschel, D.
\newblock The elliptical model of two-dimensional vortex dynamics. {I}: {T}he
 basic state.
\newblock {\mbox{\em Phys. Fluids A}} {\bf 1991}, {\em 3}, 845--854.

\bibitem[Dritschel and Legras(1991)]{dritschel91-pfa}
Dritschel, D.; Legras, B.
\newblock The elliptical model of two-dimensional vortex dynamics. {II}:
 {D}isturbance equations.
\newblock {\em Phys. Fluids A} {\bf 1991}, {\em 3},~855--869.

\bibitem[Meacham \em{et~al.}(1997)Meacham, Morrison, and Flierl]{meacham97-pf}
Meacham, S.P.; Morrison, P.J.; Flierl, G.R.
\newblock Hamiltonian moment reduction for describing vortices in shear.
\newblock {\em Phys. Fluids} {\bf 1997}, {\em 9},~2310--2328.

\bibitem[Meacham(1992)]{meacham92-dao}
Meacham, S.P.
\newblock Quasigeostrophic, ellipsoidal vortices in a stratified fluid.
\newblock {\em Dyn. Atmos. Oceans} {\bf 1992}, {\em 16}, 189--223.

\bibitem[McKiver and Dritschel(2003)]{mckiver03-jfm}
McKiver, W.J.; Dritschel, D.G.
\newblock The motion of a fluid ellipsoid in a general linear background flow.
\newblock {\mbox{\em J. Fluid Mech.}} {\bf 2003}, {\em 474},~147--173.

\bibitem[McKiver and Dritschel(2006)]{mckiver06-jfm}
McKiver, W.J.; Dritschel, D.G.
\newblock The stability of a quasi-geostrophic ellipsoidal vortex in a
 background shear flow.
\newblock {\em J. Fluid Mech.} {\bf 2006}, {\em 560},~1--17.

\bibitem[McKiver(2015)]{mckiver15-amp}
McKiver, W.J.
\newblock The ellipsoidal vortex: A novel approach to geophysical turbulence.
\newblock {\em Adv. Math. Phys.} {\bf 2015}, {\em 2015}, 613683.

\bibitem[Dritschel(2011)]{dritschel11-gafd}
Dritschel, D.G.
\newblock An exact steadily rotating surface quasi-geostrophic elliptical
 vortex.
\newblock {\em Geophys. Astrophys. Fluid Dyn.} {\bf 2011}, {\em 105},~368--376.

\bibitem[Cushman-Roisin \em{et~al.}(1985)Cushman-Roisin, Heil, and
 Nof]{cushman-roisin85-jgr}
Cushman-Roisin, B.; Heil, W.; Nof, D.
\newblock Oscillations and rotations of elliptical warm-core rings.
\newblock {\mbox{\em J. Geophys. Res.}} {\bf 1985}, {\em 20},~11756--11764.

\bibitem[Young(1986)]{young86-jfm}
Young, W.R.
\newblock Elliptical vortices in shallow water.
\newblock {\em J. Fluid Mech.} {\bf 1986}, {\em 171},~101--119.

\bibitem[Cushman-Roisin(1987)]{cushman-roisin87-tellus}
Cushman-Roisin, B.
\newblock Exact analytical solutions for elliptical vortices of the
 shallow-water equations.
\newblock {\em Tellus} {\bf 1987}, {\em 39},~235--244.

\bibitem[A.~D.~Kirwan and Liu(1991)]{kirwan91-ntop}
Kirwan, A.D., Jr.; Liu, J.
\newblock The shallow-water equations on an {F}-plane.
\newblock In {Nonlinear Topics in Ocean Physics}; Osborne, A.R., Ed.; Italian
 Physical Society: North-Holland, The Netherlands, 1991; pp. 99--132.

\bibitem[Rogers(1989)]{rogers89-pla}
Rogers, C.
\newblock Elliptic warm-core theory: The pulsrodon.
\newblock {\em Phys. Lett. A} {\bf 1989}, {\em 138},~267--273.

\bibitem[Holm(1991)]{holm91-jfm}
Holm, D.D.
\newblock Elliptical vortices and integrable {H}amiltonian dynamics of the
 rotating shallow-water equations.
\newblock {\em J. Fluid Mech.} {\bf 1991}, {\em 227},~393--406.

\bibitem[Ball(1963)]{ball63-jfm}
Ball, F.
\newblock Some general theorems concerning the finite motion of a shallow
 rotating liquid lying on a paraboloid.
\newblock {\em J. Fluid Mech.} {\bf 1963}, {\em 17},~240--256.

\bibitem[Arai and Yamagata(1994)]{arai94-chaos}
Arai, M.; Yamagata, T.
\newblock Asymmetric evolution of eddies in rotating shallow water.
\newblock {\em Chaos} {\bf 1994}, {\em 4},~163--175.

\bibitem[Stegner and Dritschel(2000)]{stegner00-jpo}
Stegner, A.; Dritschel, D.G.
\newblock A numerical investigation of the stability of isolated shallow-water
 vortices.
\newblock {\mbox{\em J. Phys. Oceanogr.}} {\bf 2000}, {\em 30},~2562--2573.

\bibitem[Chelton \em{et~al.}(2011)Chelton, Schlax, and Samelson]{chelton11-pio}
Chelton, D.B.; Schlax, M.G.; Samelson, R.M.
\newblock Global observations of nonlinear mesoscale eddies.
\newblock {\em Prog.~Oceanogr.} {\bf 2011}, {\em 91},~167--216.

\bibitem[Early \em{et~al.}(2011)Early, Samelson, and Chelton]{early11-jpo}
Early, J.J.; Samelson, R.M.; Chelton, D.B.
\newblock The evolution and propagation of quasigeostrophic ocean eddies.
\newblock {\mbox{\em J. Phys. Oceanogr.}} {\bf 2011}, {\em 41},~1535--1555.

\bibitem[Lilly \em{et~al.}(2011)Lilly, Scott, and Olhede]{lilly11-grl}
Lilly, J.M.; Scott, R.K.; Olhede, S.C.
\newblock Extracting waves and vortices from {L}agrangian trajectories.
\newblock {\mbox{\em Geophys. Res. Lett.}} {\bf 2011}, {\em 38}, doi:10.1029/2011GL049727.

\bibitem[Waterman and Lilly(2015)]{waterman15-jpo}
Waterman, S.; Lilly, J.M.
\newblock Geometric decomposition of eddy feedbacks in barotropic systems.
\newblock {\em J. Phys. Oceanogr.} {\bf 2015}, {\em 45},~1009--1024.

\bibitem[Anstey and Zanna(2017)]{anstey17-om}
Anstey, J.A.; Zanna, L.
\newblock A deformation-based parametrization of ocean mesoscale eddy
 {R}eynolds stresses.
\newblock {\em Ocean Model.} {\bf 2017}, {\em 112},~99--111.

\bibitem[Lilly and Olhede(2010)]{lilly10-itsp}
Lilly, J.M.; Olhede, S.C.
\newblock Bivariate instantaneous frequency and bandwidth.
\newblock {\em IEEE Trans. Signal Process.} {\bf 2010}, {\em 58},~591--603.

\bibitem[Lilly and Gascard(2006)]{lilly06-npg}
Lilly, J.M.; Gascard, J.C.
\newblock Wavelet ridge diagnosis of time-varying elliptical signals with
 application to an~oceanic eddy.
\newblock {\em Nonlinear Process. Geophys.} {\bf 2006}, {\em 13},~467--483.

\bibitem[Beron-Vera \em{et~al.}(2008)Beron-Vera, Olascoaga, and
 Goni]{beron-vera08-grl}
Beron-Vera, F.J.; Olascoaga, M.J.; Goni, G.J.
\newblock Oceanic mesoscale eddies as revealed by {L}agrangian coherent
 structures.
\newblock {\em Geophys. Res. Lett.} {\bf 2008}, {\em 35},~L12603.

\bibitem[Rutherford \em{et~al.}(2012)Rutherford, Dangelmayr, and
 Montgomery]{rutherford12-acp}
Rutherford, B.; Dangelmayr, G.; Montgomery, M.T.
\newblock Lagrangian coherent structures in tropical cyclone intensification.
\newblock {\em Atmos. Chem. Phys.} {\bf 2012}, {\em 12},~5483--5507.

\bibitem[Haller and Beron-Vera(2012)]{haller12-pd}
Haller, G.; Beron-Vera, F.J.
\newblock Geodesic theory of transport barriers in two-dimensional flows.
\newblock {\em Phys. D} {\bf 2012}, {\em 241},~1680--1702.

\end{thebibliography}


\end{document}